\documentclass[submission, Phys]{SciPost}

\graphicspath{{plots/}}
\usepackage{subcaption}
\usepackage{graphicx}
\usepackage{caption}
\usepackage{subcaption}

\newcommand{\xph}{{ \overline{ph} }}

\newcommand{\M}{{ \mathrm{M} }}
\newcommand{\D}{{ \mathrm{D} }}
\newcommand{\SC}{{ \mathrm{SC} }}

\usepackage[noabbrev]{cleveref}
\newcommand{\ceq}[1]{Eq.~\eqref{eq:#1}}
\newcommand{\cfg}[1]{Fig.~\ref{fig:#1}}

\hypersetup{
    colorlinks,
    linkcolor={red!50!black},
    citecolor={blue!50!black},
    urlcolor={blue!80!black}
}

\usepackage[bitstream-charter]{mathdesign}
\DeclareSymbolFont{usualmathcal}{OMS}{cmsy}{m}{n}
\DeclareSymbolFontAlphabet{\mathcal}{usualmathcal}

\begin{document}

\begin{center}{\Large \textbf{
Entangled magnetic, charge, and superconducting pairing correlations in the two-dimensional Hubbard model: a functional renormalization-group analysis\\
}}\end{center}

\begin{center}
Sarah Heinzelmann\textsuperscript{1$\star$},
Alessandro Toschi\textsuperscript{2}, and
Sabine Andergassen\textsuperscript{1,2,3}
\end{center}

\begin{center}
{\bf 1} Institut f\"ur Theoretische Physik and Center for Quantum Science, Universit\"at T\"ubingen,
T\"ubingen, Germany
\\
{\bf 2} Institute for Solid State Physics, Vienna University of Technology,
Vienna, Austria
\\
{\bf 3} Institute of Information Systems Engineering, Vienna University of Technology,
Vienna, Austria
\\
${}^\star$ {\small \sf sarah.heinzelmann@uni-tuebingen.de}
\end{center}

\begin{center}
\today
\end{center}

\section*{Abstract}
{\bf
Using the recently introduced multiloop extension of the functional renormalization group, we compute the magnetic, density, and superconducting susceptibilities of the two-dimensional Hubbard model at weak coupling and present a detailed analysis of their evolution with temperature, interaction strength, and loop order.
By breaking down the susceptibilities into contributions from the bare susceptibility and the individual channels, we investigate their relative importance as well as the channel interplay. 
In particular, we trace the influence of antiferromagnetic fluctuations on the $d$-wave superconductivity and provide an analytical understanding for the observed behavior. 
}

\vspace{10pt}
\noindent\rule{\textwidth}{1pt}
\tableofcontents\thispagestyle{fancy}
\noindent\rule{\textwidth}{1pt}
\vspace{10pt}

\section{Introduction}
\label{sec:introduction}

The ground state phase diagram of the two-dimensional (2D) Hubbard model at weak coupling reveals a rich behavior as a function of the different parameters (see Ref.~\citeonline{Qin2021} for a recent overview of computational results). 
In addition to antiferromagnetic (AF) N\'{e}el order, magnetic order with generally incommensurate wave vectors away from the N\'eel point $(\pi,\pi)$ can be found away from half filling in mean-field studies \cite{Schulz1990,Dombre1990,Fresard1991,Igoshev2010} and, including fluctuations, by expansions in the hole-density \cite{Shraiman1989,Chubukov1992,Chubukov1995,Kotov2004}.
Incommensurate magnetic order is also indicated by diverging interactions and susceptibilities 
in functional renormalization group (fRG) flows \cite{Halboth2000b,Husemann2009,Metzner2012}, where approximate solutions indicate robust magnetic order up to fairly high doping provided that superconductivity is suppressed \cite{Yamase2016}.

While the magnetic instability in the Hubbard model is reproduced already by conventional mean-field theory, pairing is fluctuation-driven and hence more difficult to capture. Simple qualitative arguments suggesting $d$-wave pairing driven by magnetic fluctuations were corroborated by the fluctuation exchange approximation \cite{Scalapino1995}.
Convincing evidence for superconductivity at weak and moderate coupling strengths has been established by self-consistent or renormalized perturbation expansions \cite{Bickers1989,Neumayr2003,Kyung2003,Raghu2010} and from fRG calculations \cite{Zanchi2000,Halboth2000b,Honerkamp2001a,Honerkamp2001b,Metzner2012,Eberlein2014}.
With its unbiased treatment of all fluctuation channels on equal footing, the fRG confirmed earlier studies based on the summation of certain perturbative contributions \cite{Scalapino2012},
finding $d$-wave superconductivity with a sizable gap already. 

In the fRG, a symmetry breaking such as magnetic order or superconductivity is signaled by a divergence of the two-particle vertex at a finite scale $\Lambda_c > 0$.
To continue the flow beyond this critical scale, 
an infinitesimally small symmetry breaking field has to be added to the bare action, which develops into a finite order parameter below $\Lambda_c$ \cite{Salmhofer2004}. Within the one-loop ($1\ell$) truncation, 
the two-particle vertex function always diverges in some momentum channels 
(even for a small bare interaction) 
and 
one always runs into a strong coupling problem in the low-energy limit.
The $1\ell$ truncation breaks down, and also simplified parametrizations of the two-particle vertex cannot be justified in the presence of singular momentum and energy dependencies.
From the stability analysis at 
sufficiently small $U$ (and away from the van Hove singularity\footnote{When the chemical potential approaches the van Hove singularity, different instabilities compete, involving also ferromagnetism 
\cite{Irkhin2001,Hlubina1997,Honerkamp2001b,Katanin2003,Neumayr2003,Raghu2010}.}), 
superconductivity with a $d$-wave order parameter can be inferred from the divergence of the 
pairing susceptibility \cite{Metzner2012}.
The combination of the flow equations at high scales with a mean-field approximation at low scales 
showed that a sizable next-nearest neighbor hopping amplitude favors $d$-wave superconductivity \cite{Eberlein2014} and revealed a robust $d$-wave pairing coexisting with N\'eel or incommensurate antiferromagnetism \cite{Reiss2007,Wang14,Yamase2016}.
At finite temperature, the off-diagonal long-range order turns into the quasi long-range order of a Kosterlitz-Thouless phase.
Computing the Kosterlitz-Thouless transition temperature from the superfluid phase stiffness, a superconducting dome centered around optimal doping has been found \cite{Vilardi2020}.
The interplay of antiferromagnetism and superconductivity in the Hubbard model has also been analyzed by fRG flows with order-parameter fields. It was clarified how $d$-wave pairing mediated by AF fluctuations emerges from the coupled fermion-boson flow \cite{Krahl2009}, and a phase diagram with magnetic and superconducting order was computed at weak coupling \cite{Friederich2010,Friederich2011}.
More recently, fRG flows starting from the dynamical mean-field theory (DMFT) \cite{Metzner1989,Georges1996} solution instead of the bare action \cite{Taranto2014}
confirmed robust pairing with $d$-wave symmetry also at strong coupling, driven by magnetic correlations at the edge of the AF regime \cite{Vilardi2019,Bonetti2021}.

In this work, we use a forefront algorithmic implementation of the multiloop extension of the 
fRG \cite{Tagliavini2019,Hille2020}, which 
includes an accurate treatment of both the frequency and momentum dependencies of the two-particle vertex 
and allows to sum up all the diagrams of the parquet approximation with their exact weight \cite{Kugler2018a,Kugler2018b}, yielding cutoff-independent results \cite{Tagliavini2019,Chalupa2022}.
The application to the 2D Hubbard model both at half filling and finite doping presented here focuses on the analysis of the different fluctuation channels and their impact on the observed physical behavior.

The paper is organized as follows: In Section~\ref{sec:modelandmethod} we introduce the Hubbard model and briefly review the fRG implementation used for the present study.
In Section~\ref{sec:results} we discuss the results for the susceptibilities in the different channels, together with a fluctuation diagnostics that allows to identify the relevant degrees of freedom driving the physical behavior in the different parameter regimes. 
We finally conclude with a summary and an outlook in Section~\ref{sec:conclusions}.

\section{Model and method}
\label{sec:modelandmethod}

\subsection{Two-dimensional Hubbard model}
\label{sec:model}

We consider the single-band Hubbard model in two dimensions, defined by the Hamiltonian
\begin{align}
\label{eq:defhamilt}
\hat{\mathcal{H}}=\sum_{i,j, \sigma}t_{ij}\hat{c}^{\dagger}_{i\sigma}\hat{c}_{j\sigma}+U\sum_i \hat{n}_{i\uparrow}\hat{n}_{i\downarrow} - \mu \sum_{i,\sigma}\hat{n}_{i\sigma} \;,
\end{align}
where $\hat{c}_{i\sigma}$ ($\hat{c}^{\dagger}_{i\sigma}$) annihilates (creates) an electron with spin $\sigma$ at the lattice site $i$ and $\hat{n}_{i\sigma}=\hat{c}^{\dagger}_{i\sigma}\hat{c}_{i\sigma}$ is the spin-resolved number operator, with half filling $n=\left<n_\uparrow + n_\downarrow\right>=1$ corresponding to a chemical potential of $\mu=U/2$. $t_{ij}=-t$ describes the hopping amplitude between nearest neighbors and $t_{ij}=-t'$ the one between next-nearest neighbors. Finally, $U$ is the strength of the (purely local) Coulomb interaction.

The bare propagator is
\begin{align} \label{eq:G0}
    G_0({\bf k},i\nu)= \big( i\nu +\mu -\epsilon_{\bf k} \big)^{-1} \;,
\end{align}
with
\begin{align}
    \epsilon_{\bf k}=-2t( \cos{k_x} + \cos{k_y} ) - 4t'\cos{k_x}\cos{k_y} \;.
\end{align}
For $t'=0$, the dispersion exhibits a 
diamond-shaped Fermi surface at half filling, resulting in an interesting behavior already for the non-interacting case: the perfect nesting for 
the momentum vector $\mathbf{q}=(\pi,\pi)$ 
leads to an enhanced susceptibility at $\mathbf{q}$. 

In the following, we use $t\equiv 1$ as the energy unit. Furthermore, we set $\hbar=1$ and $k_\text{B}=1$.

\subsection{Functional renormalization group}
\label{frg}

Over the last 
decades, fRG-based approaches have been broadly used for analyzing 2D lattice electron systems, 
see Refs.~\cite{Salmhofer1999,Berges2002,Kopietz2010,Metzner2012,Dupuis2021} for a review, and 
more recently also spin systems \cite{Mueller2023}.
Their characteristic scale-dependent behavior 
can be treated in a flexible and unbiased way by the fRG. 
Its starting point is an exact functional flow equation, which yields the gradual evolution from a microscopic model action to the final effective action as a function of a continuously decreasing energy scale. Expanding in powers of the fields one obtains an exact hierarchy of flow equations for the $n$-particle irreducible vertex functions. In practical implementations, this infinite hierarchy is truncated at the two-particle level: neglecting the renormalization of three- and higher order particle vertices yields approximate $1\ell$ flow equations for the self-energy and two-particle vertex. 
The underlying approximations are devised for the weak to intermediate coupling regime\footnote{More strongly correlated parameter regimes 
become accessible by exploiting correlated 
starting points for the fRG flow \cite{Wentzell2015}.}, where forefront algorithmic advancements brought the fRG 
to a quantitatively reliable level \cite{Hille2020}.
The substantial improvement with respect to previous fRG-based computation schemes relies on an efficient parametrization of the two-particle vertex which takes into account both the momentum and frequency dependence. 
The comparison to determinant Quantum Monte Carlo data has shown in fact, that the multiloop fRG is remarkably accurate up to moderate interaction strengths \cite{Hille2020}.

Here, however, we do not aim at a quantitative analysis, but rather at a fundamental description of the physical behavior. For this reason, we use 
the $2\ell$ approximation as a first step towards the multiloop fRG, with a substantially reduced numerical effort with respect to fully loop converged computations. We retain only the multiloop equation for the self-energy, which, differently to the conventional $1\ell$ flow,  accounts for the pseudogap opening \cite{Hille2020b}.
In the following, we briefly introduce the formalism. 

Assuming translational invariance with $U(1)$ charge and $SU(2)$ spin-rotation symmetry, the one-particle irreducible vertex can be expressed 
by a coupling function depending on three independent generalized momenta 
via \cite{Honerkamp2001a}
\begin{align}
V(k_1,k_2,k_3,k_4)_{\sigma_1,\sigma_2,\sigma_3,\sigma_4}=& -\delta_{\sigma_1,\sigma_4}\delta_{\sigma_2,\sigma_3} V(k_1,k_4,k_3)+ \delta_{\sigma_1,\sigma_2}\delta_{\sigma_3,\sigma_4}  V(k_1,k_2,k_3) \; ,
\end{align}
with $k_4=k_1+k_3-k_2$ due to energy and momentum conservation. $\sigma_i$ represents the spin quantum number and $k_i=(\mathbf{k_i},\nu_i)$ includes both the momentum and Matsubara frequency. 
The coupling function can be further decomposed into the magnetic, density, and superconducting channel by the parquet decomposition 
\cite{Dedominicis1964,Dedominicis1964A,Bickers2004,Yang2009,Tam2013,Rohringer2012,Valli2015,Kauch2019,Li2019}
\begin{align}
\label{eq:parquet}
    V(k_1,k_2,k_3)=&\,\Lambda^{\mathrm{2PI}}-\frac{1}{2}\Phi^{\M}(k_2-k_1,k_1,k_4)
    -\Phi^{\M}(k_3-k_2,k_1,k_2) \nonumber \\
    &+\frac{1}{2}\Phi^{\D}(k_2-k_1,k_1,k_4)+\Phi^{\SC}(k_1+k_3,k_1,k_2)\; .
\end{align}
The relations between the physical channels and their diagrammatic counterparts are outlined in Appendix \ref{app:notation}.
In the parquet approximation, the fully two-particle irreducible vertex is approximated by $\Lambda^{\mathrm{2PI}}=U$
and the two-particle reducible contributions $\Phi^{\eta}$, where $\eta=\M/\D/\SC$ indicates the magnetic (or spin), density, and superconducting channel,
are parametrized using a single generalized bosonic transfer momentum/frequency as first argument and two fermionic ones as second and third argument.
In particular, we use a form-factor expansion for 
the fermionic momentum dependence within the so-called "truncated unity" or TU-fRG \cite{Husemann2009,Wang2012,Lichtenstein2017}. For the frequencies, we take into account the full dependence at low frequencies (see also Ref. \cite{Vilardi2017}), supplemented by 
the fermionic high-frequency asymptotics \cite{Rohringer2012,Wentzell2016}. 
The explicit flow equations, together with the technical parameters employed for their 
numerical implementation, are reported in Appendix \ref{app:impl}.

\subsection{Computation 
of the susceptibilities and fluctuation diagnostics}

We compute the physical susceptibilities 
$\chi^{\eta}(\mathbf{q},i\omega)$ in Matsubara frequency space.
For their definition and the employed conventions we refer to Ref. \cite{Hille2020}.

As for the 
$n$-particle vertex functions, differential flow equations can be derived also for the response functions and susceptibilities
\cite{Tagliavini2019}. 
Alternatively, the susceptibilities 
can be determined by a so-called "post-processing" procedure, where only the self-energy and two-particle vertex at the end of the flow are inserted into the exact equation
\begin{align}
\label{eq:susc_postproc}
    \mathbf{\chi}^{\eta}(\mathbf{q},i\omega)
    &= \sum_{i\nu}\mathbf{\Pi}^{\eta}(\mathbf{q},i\omega,i\nu) +  
    \sum\limits_{\substack{i\nu,i\nu'}} \mathbf{\Pi}^{\eta}(\mathbf{q},i\omega,i\nu) \mathbf{V}^{\eta}(\mathbf{q},i\omega,i\nu,i\nu') \mathbf{\Pi}^{\eta}(\mathbf{q},i\omega,i\nu') \nonumber\\
    &:= \chi^{\eta,0}(\mathbf{q},i\omega) + \left[\chi^{\eta,0}\mathbf{V}^{\eta}\chi^{\eta,0}\right](\mathbf{q},i\omega)  \;,
\end{align}
where the definitions of $\mathbf{V}^\eta$ and the relations between quantities in different channels can be found in Appendix \ref{app:notation}.
Since bold symbols represent matrices in form-factor space, Eq.~\eqref{eq:susc_postproc} is to be read as including a matrix multiplication.
The components 
of $\mathbf{\Pi}^\eta$ in the form-factor basis are obtained by 
\begin{align}
\label{eq:pi}
   \mathbf{\chi}^{\eta,0}_{mn}(\mathbf{q},i\omega)&\left(=\sum_{i\nu}\Pi^{\eta}_{mn}(\mathbf{q},i\omega,i\nu) \right)
   \nonumber \\
   &=\text{sgn}(\eta)\sum_{i\nu}\int \! \text{d}\mathbf{p}f^*_n(\mathbf{p})f_m(\mathbf{p})G(\mathbf{p},i\nu)G(\mathbf{q}-\text{sgn}(\eta)\mathbf{p},i\omega -\text{sgn}(\eta) i\nu)\;,
\end{align}
with $f_{0}(\mathbf{k})=1$ for the $s$- and $f_{1}(\mathbf{k})=\cos(k_x)-\cos(k_y)$ for the $d$-wave form factor and 
sgn($\eta$)=$+/-$
for $\eta=\SC,\D/\M$ respectively.
The compact representation introduced above 
for the matrix product including the sum over the fermionic frequencies highlights the structure of the two contributions to the susceptibility: the bare bubble and the vertex correction.
These can be determined individually. Moreover, 
\ceq{susc_postproc} represents the starting point 
for the diagnostics of the prevailing fluctuations \cite{Gunnarsson2015,Rohringer2020rev,Schaefer2021b,Delre2021,Dong2022}, where the vertex correction is further broken down into separate contributions 
from the bare vertex and the different channels by using the parquet decomposition of \ceq{parquet}. In the following, we will focus on the static susceptibilities at zero frequency $(i\omega=0)$.

We note that in the multiloop fRG, both computation schemes, from the flow of the response functions and from post-processing, become equivalent at loop convergence \cite{Tagliavini2019,Kugler2018c}. 
At finite loop order, we use 
the post-processing where the multiloop corrections affect the 
susceptibilities only at $\mathcal{O} (U^3)$ (the flowing ones 
at $\mathcal{O} (U^2)$) in the renormalized interaction and hence converge faster in the number of loops \cite{Hille2020}. 

\section{Results} 
\label{sec:results}

\begin{figure}
    \centering
    \includegraphics[width=0.9\linewidth]{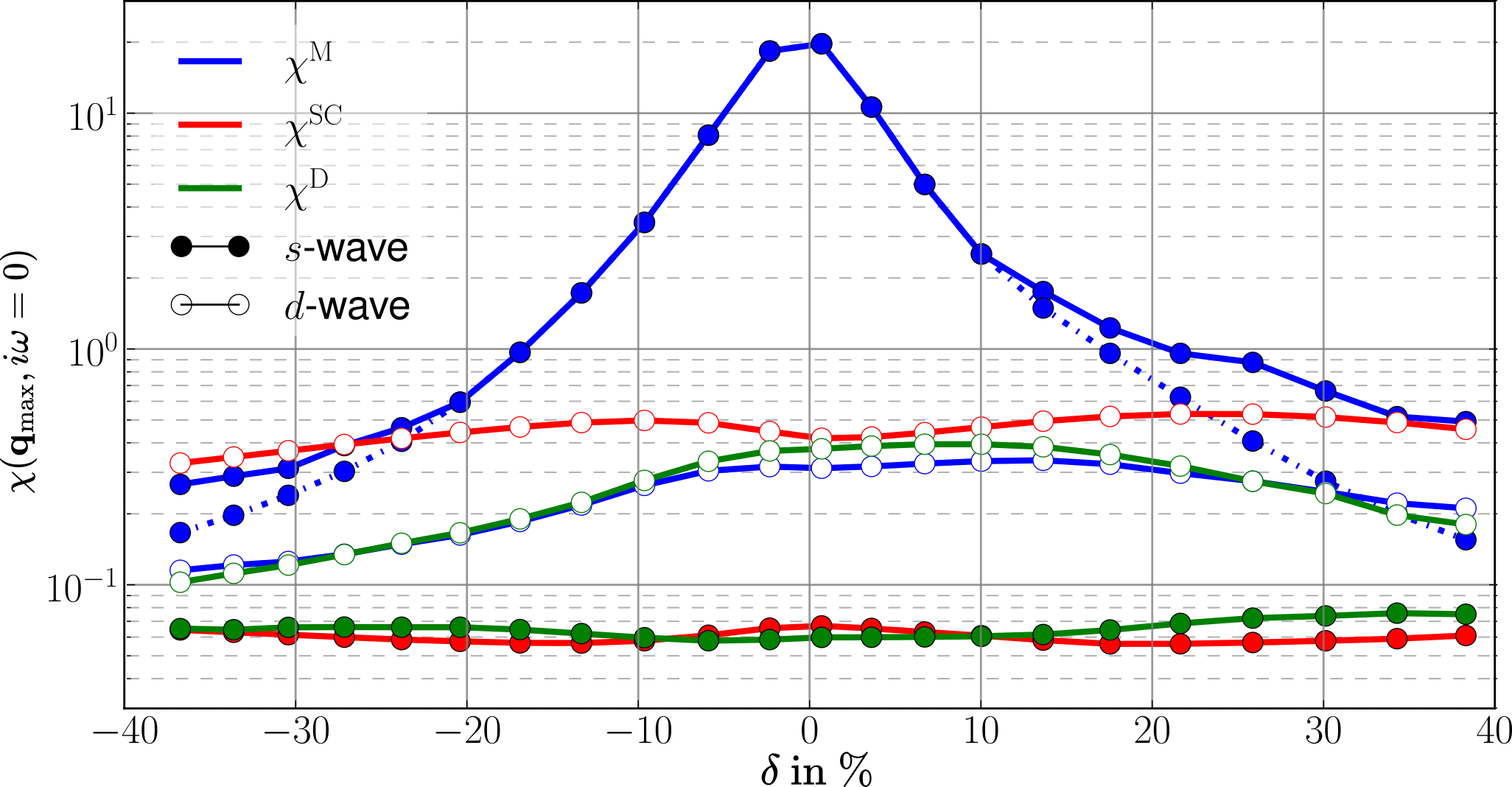}
    \caption{Maximal values of $s$- and $d$-wave magnetic $\chi^{\M}(\mathbf{q}_\text{max},i\omega=0)$, density $\chi^{\D}(\mathbf{q}_\text{max},i\omega=0)$, and superconducting $\chi^{\SC}(\mathbf{q}_\text{max},i\omega=0)$ susceptibilities as obtained from the fRG in the $2\ell$ truncation (by post-processing), 
    for $U=3$, $t'=-0.15$, and $\beta=15$.
    We refer to the text for the definition of $\mathbf{q}_\text{max}$.
    In the magnetic channel, the value at $\mathbf{q}=(\pi,\pi)$ is added (dashed line). Note the logarithmic scale, highlighting the different orders of magnitude.} 
    \label{fig:phasediagram_overview}
\end{figure}

\begin{figure}
    \centering
    \includegraphics[width=1.\linewidth]{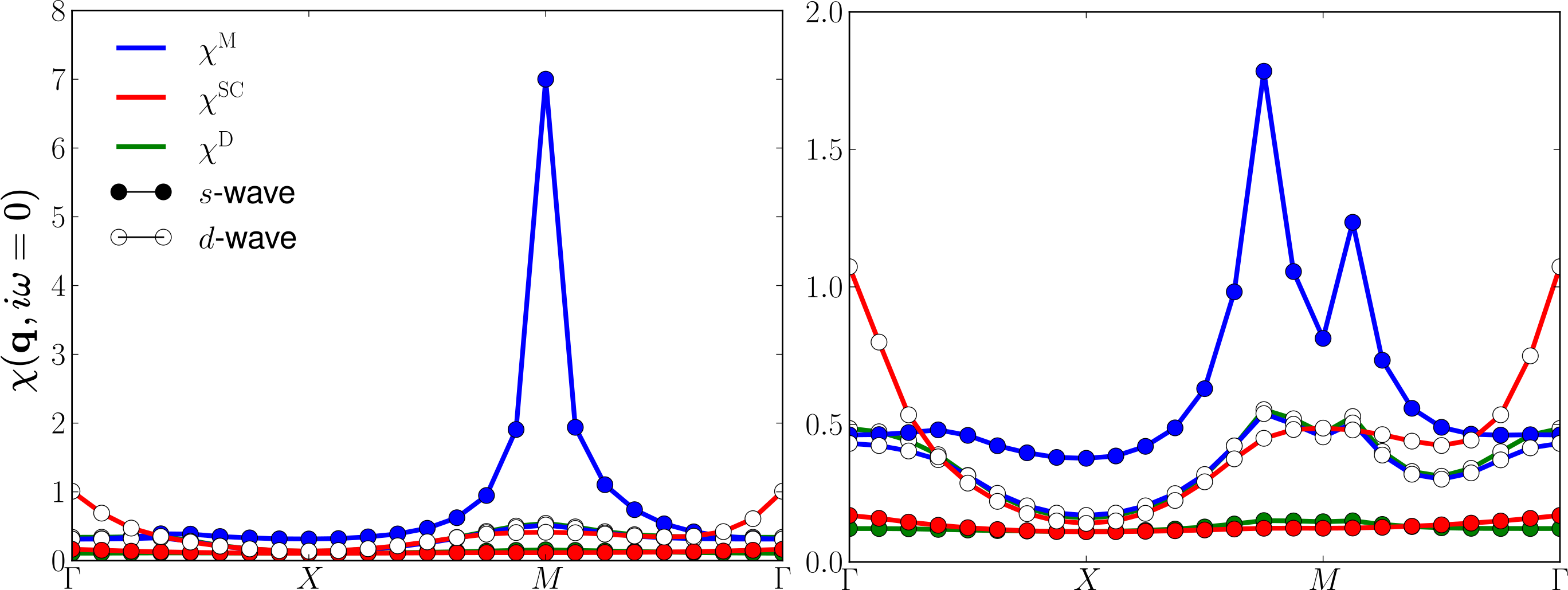}
    \caption{Susceptibilities of Fig. \ref{fig:phasediagram_overview} along a path in the Brillouin zone, for $\delta=10~\%$ electron and $26~\%$ hole doping on the left and right panel respectively. The bubble and vertex contributions are reported in Appendix \ref{app:altfig2}.}
    \label{fig:susc_alongBZ}
\end{figure}

An overview of the $2\ell$ results for the static magnetic, density, and superconducting susceptibilities for the representative parameters of $U=3$, $t'=-0.15$, and the inverse temperature $\beta=15$ is shown 
in Fig.~\ref{fig:phasediagram_overview}, reporting the evolution of their maximal values 
with the doping $\delta=(1-n)$,  
where $n$ is the occupation number. 
The asymmetry around half filling is a consequence of the finite next-nearest neighbor hopping amplitude $t'$ breaking particle-hole symmetry. 
The dominant susceptibility is the $s$-wave magnetic $\chi^\M$ one, for almost all dopings except for very large 
values, where two-particle interaction effects play a vanishing role and the physics is dominated by Fermi-liquid behavior. 
The maximum around half filling occurs in correspondence of the 
AF wave vector $\mathbf{q}=(\pi,\pi)$. Its height decreases with the doping leading to a damping at large dopings. 
At the same time, $\chi^\M$ eventually becomes incommensurate at around $10\%$ hole and $20\%$ electron doping, as indicated by the departure of the solid from the dashed line (see also Fig. \ref{fig:susc_alongBZ} discussed below).
The largest subleading susceptibility is the
$d$-wave superconducting $\chi^{\SC}$ one, followed by the other $d$-wave components $\chi^{\D}$ and $\chi^{\M}$, while the $s$-wave $\chi^{\D}$ and $\chi^{\SC}$ appear to be 
much smaller.
The considered parameter regime is still far away from any instability and we expect the $d$-wave pairing susceptibility to increase only at lower temperatures, see also discussion below. 

The characteristic behavior along a path in the Brillouin zone is illustrated in Fig.~\ref{fig:susc_alongBZ} (see also Appendix \ref{app:altfig2} for the bubble and vertex contributions), both around half filling and at finite doping.
Close to half filling, shown in the left panel, the dominant feature is the AF peak of $\chi^{\M}$ at $\mathbf{q}=(\pi,\pi)$. The situation is more interesting at finite doping, reported in the right panel: the AF peak splits and shifts to incommensurate wave vectors. At the same time, the $d$-wave 
$\chi^{\SC}$ becomes comparable in size to the magnetic one at its maximum value in correspondence of $\mathbf{q}=(0,0)$. 
In the following analysis $\mathbf{q}_\text{max}$ refers to the wave vector 
where the susceptibility reaches its maximum. We note that it depends on the channel and doping.
For completeness, the evolution of the Fermi surface as extracted from the self-energy at the lowest Matsubara frequency, is provided in Appendix \ref{app:fermisurface}.

\begin{figure}
    \centering
    \includegraphics[width=0.9\linewidth]{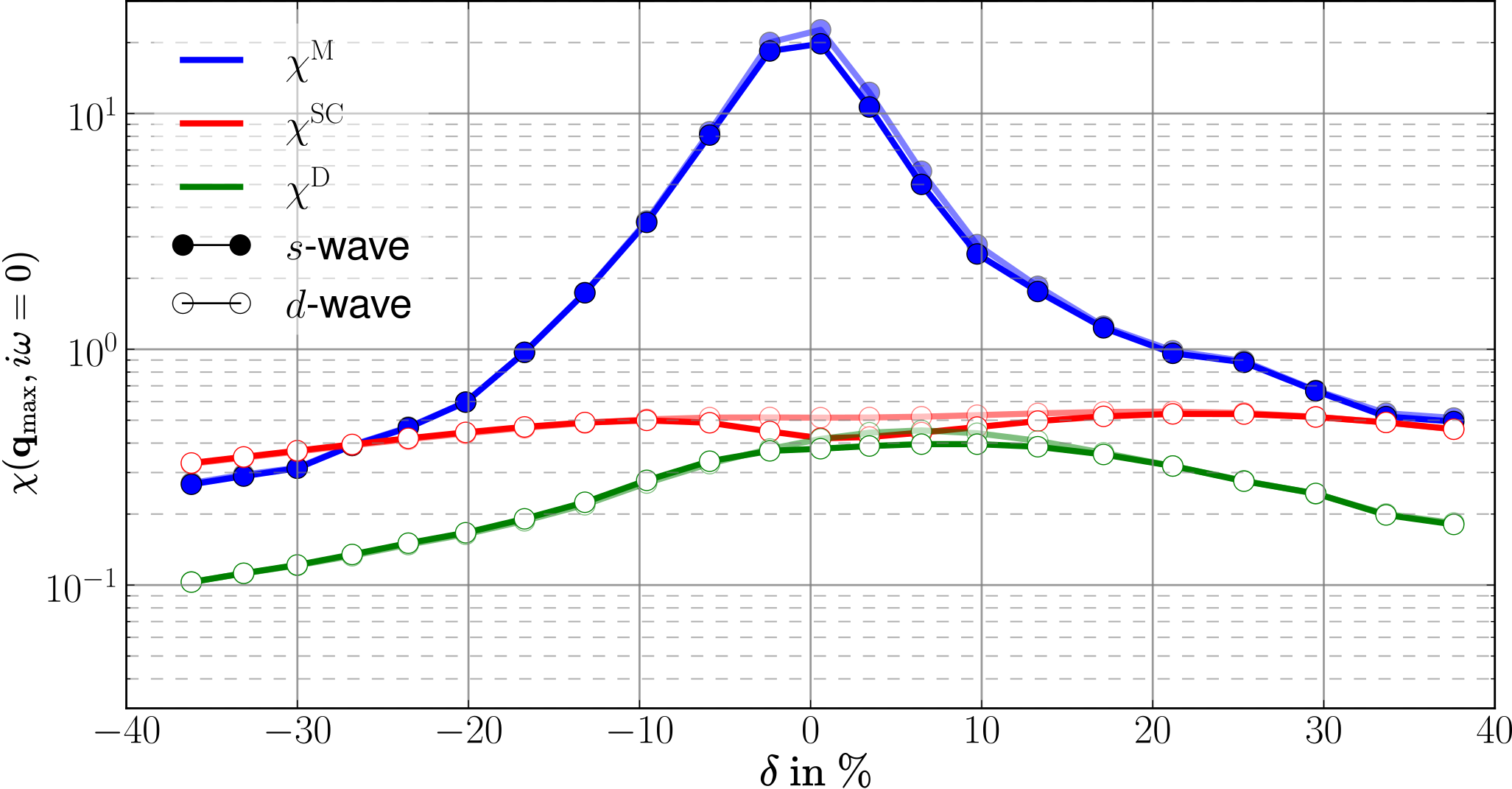}
    \caption{Leading 
    magnetic $\chi^{\M}(\mathbf{q}_{\text{max}},i\omega=0)$,  
    density $\chi^{\D}(\mathbf{q}_{\text{max}},i\omega=0)$, and superconducting $\chi^{\SC}(\mathbf{q}_{\text{max}},i\omega=0)$ susceptibilities of Fig.~\ref{fig:phasediagram_overview}, with comparison of the results obtained from the flow (lighter colors) to the ones from post-processing (darker colors).     }
    \label{fig:phasediagram_overview1}
\end{figure}

While the $2\ell$ computations are not converged in loop order, they include the first
explicit correction terms to the two-particle vertex flow \cite{Katanin2009,Eberlein2014a} extending the Katanin substitution. On a qualitative level, the $2\ell$ truncation is
correct up to $O(U^3)$, 
in contrast to $O(U^2)$ of the conventional $1\ell$ scheme.
On a quantitative level, the $2\ell$ contributions 
are typically the most important ones \cite{Tagliavini2019,Hille2020}. 
This is confirmed by the small deviations with respect to the results obtained from the flow of the response function, reported in Fig.~\ref{fig:phasediagram_overview1}. Except for the region around half filling, we observe almost no differences with respect to the post-processing data 
of Fig.~\ref{fig:phasediagram_overview}.
A more detailed analysis of the effects of higher loop orders is presented in Section \ref{sec:multiloop}.

In the following we will perform a diagnostics of the fluctuations by analysing the contribution of the different channels 
to identify the dominant two-particle scattering processes controlling the physical response of the system. We will 
also discuss the effects of higher loop orders and explore different parameter regimes.

\subsection{Magnetic susceptibility}
\label{sec:magnetism}

The $s$-wave component of $\chi^\M$ is the dominant 
susceptibility at half filling, exceeding the largest subleading susceptibilities by two orders of magnitudes (see Fig.~\ref{fig:phasediagram_overview}). 
The pronounced peak around the N\'eel wave vector $\bf{q}=(\pi,\pi)$ indicates strong AF fluctuations. 
At finite doping, its absolute value decreases considerably with increasing $\delta$. 
We also observe that the wave vector $\mathbf{q}_{\mathrm{max}}$ corresponding to the maximum becomes incommensurate, 
as expected for larger dopings \cite{Metzner2012,Halboth2000a,Halboth2000b,Yamase2016}. 
At hole doping, 
the associated maximal value of $\chi^\M$ exhibits a visible "shoulder". 

\begin{figure}
    \centering
    \includegraphics[width=0.9\linewidth]{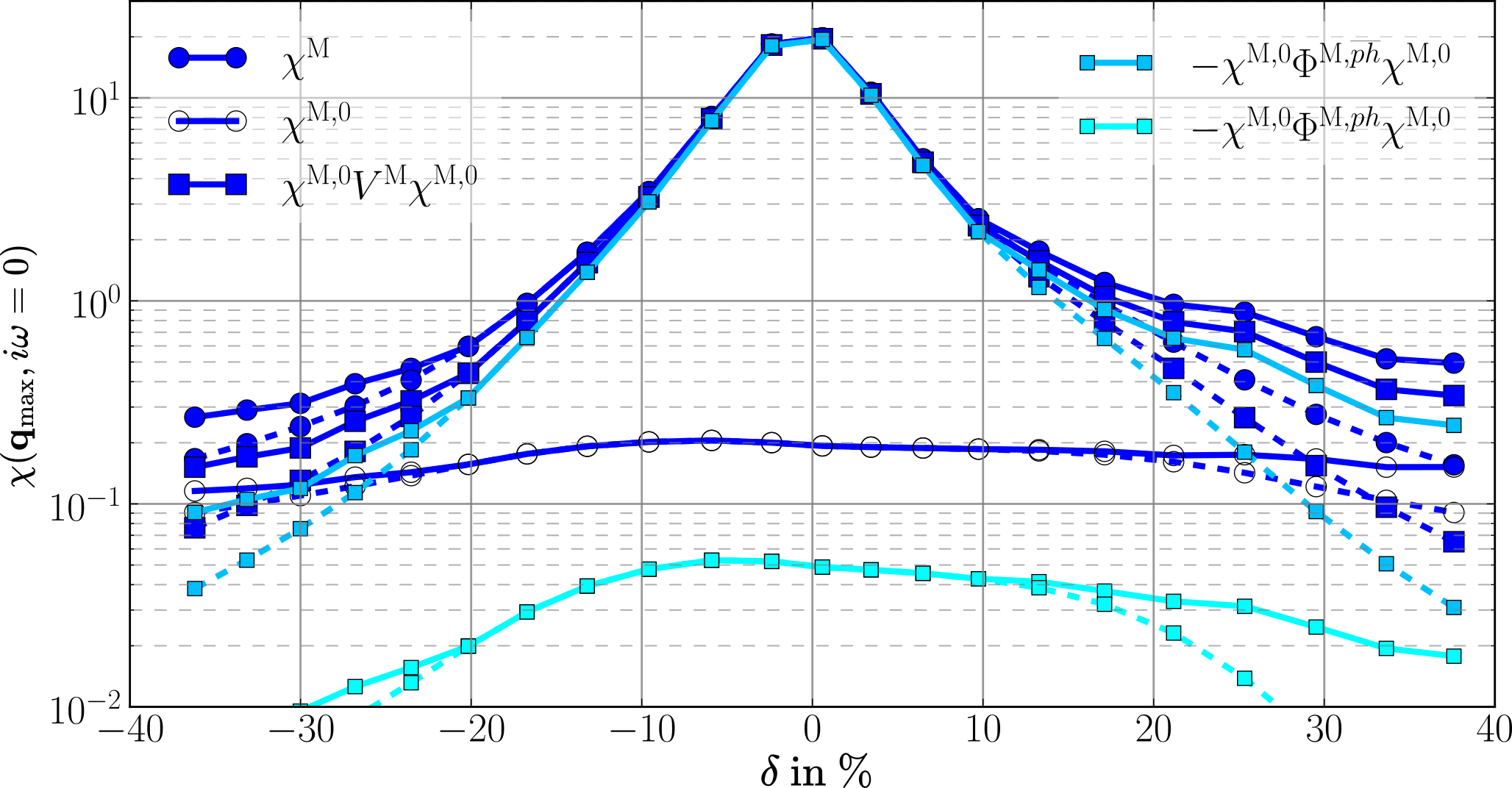}
    \caption{Diagnostics of the fluctuations for the $s$-wave magnetic susceptibility $\chi^{\M}(\mathbf{q}_{\text{max}},i\omega=0)$ 
    of Fig.~\ref{fig:phasediagram_overview},
    evaluated both in correspondence of $\mathbf{q}_\mathrm{max}$ (solid lines) and $\mathbf{q}=(\pi,\pi)$ (dashed lines).
    The vertex corrections dominate over the bare bubble and originate essentially from  
    the contribution of the crossed particle-hole channel $\chi^{\M,0}\mathbf{\Phi}^{\M,\xph}\chi^{\M,0}$, 
    whereas the one from   
    the particle-hole channel $\chi^{\M,0}\mathbf{\Phi}^{\M,ph}\chi^{\M,0}$ appears to be negligible 
    (the contributions from $\chi^{\M,0}\mathbf{\Phi}^{\SC}\chi^{\M,0}$ and 
    $\chi^{\M,0}\mathbf{\Phi}^{\D}\chi^{\M,0}$ are smaller than $0.01$ and not reported). }
    \label{fig:pp_Ms}
\end{figure}

In order to perform a fluctuation diagnostics for the $s$-wave magnetic susceptibility $\chi^\M$ 
and thereby trace the 
driving contributions to the AF peak, we insert 
the self-energy and the two-particle vertex obtained from the flow into the post-processing 
Eq.~\eqref{eq:susc_postproc} for $\eta=\M$.
Using 
the parquet decomposition 
for the magnetic vertex, yields 
\begin{align}
    \mathbf{V}^{\M}
    = &  - \mathbf{\Lambda}^{2\text{PI}}+\mathbf{\Phi}^{\M} +\frac{1}{2}P^{ph\rightarrow \xph}\left(\mathbf{\Phi}^\M
      -\mathbf{\Phi}^\D\right) 
       - P^{pp\rightarrow \xph}\mathbf{\Phi}^{\SC} 
       \;,
\end{align}
with $\Lambda^{2\text{PI}}_{mn}=-U\delta_{m,0}\delta_{n,0}$ and the projection operators 
$P^{r\rightarrow r'}$ between the diagrammatic channels $r$, $r'=pp$, $ph$, $\overline{ph}$ 
which are required for the translation of the 
channel-specific notation, see Appendix \ref{app:notation}.  
As a consequence, there are two contributions of $\mathbf{\Phi}^\M$:  
one originating from $\mathbf{\Phi}^\xph=-\mathbf{\Phi}^\M$ and the other one from in $\mathbf{\Phi}^{ph}=-\left(\mathbf{\Phi}^\M-\mathbf{\Phi}^\D\right)/2$, in which the 
projection from the particle-hole 
to the crossed particle-hole 
notation used for $\mathbf{V}^\M$  
leads to  
a different momentum and frequency dependence.
Thus, the magnetic susceptibility can be explicitly decomposed in 
\begin{align}
    \mathbf{\chi}^{\M}
    (\mathbf{q},i\omega) 
    = & \mathbf{\chi}^{\M,0}(\mathbf{q},i\omega)
    - \left[\mathbf{\chi}^{\M,0}\mathbf{\Lambda}^{2\text{PI}}\mathbf{\chi}^{\M,0}\right](\mathbf{q},i\omega)\nonumber\\&
    +\left[\chi^{\M,0}\mathbf{\Phi}^{\M}\chi^{\M,0}\right](\mathbf{q},i\omega)
    + \frac{1}{2}\left[\chi^{\M,0}P^{ph\rightarrow \xph}\left(\mathbf{\Phi}^\M-\mathbf{\Phi}^\D\right)\chi^{\M,0}\right](\mathbf{q},i\omega)\nonumber\\
    &
    - \left[\chi^{\M,0}P^{pp\rightarrow \xph}\mathbf{\Phi}^{\SC}\chi^{\M,0}\right](\mathbf{q},i\omega)
     \;,
     \label{eq:suscsplit_m}
\end{align}
which allows to evaluate 
the importance of the different channels. 
As shown in Fig.~\ref{fig:pp_Ms}, the static ($s$-wave) magnetic susceptibility $\chi^\M$ is mostly driven by the 
vertex correction, in particular by the
contribution of the crossed particle-hole channel  
$\mathbf{\Phi}^\M=-\mathbf{\Phi}^{\xph}$ and not by 
$P^{ph\rightarrow\xph}\mathbf{\Phi}^\M$ (in the figure we refer to 
$\mathbf{\Phi}^{\M,\xph}$ and $\mathbf{\Phi}^{\M,ph}$, respectively). 
In fact, in the projection from the particle-hole to the crossed particle-hole channel only the momentum average adds to the latter.
All other contributions appear negligible for the considered parameters of $U=3$, $t'=-0.15$, and $\beta=15$. 

\begin{figure}
    \centering
    \includegraphics[width=0.9\linewidth]{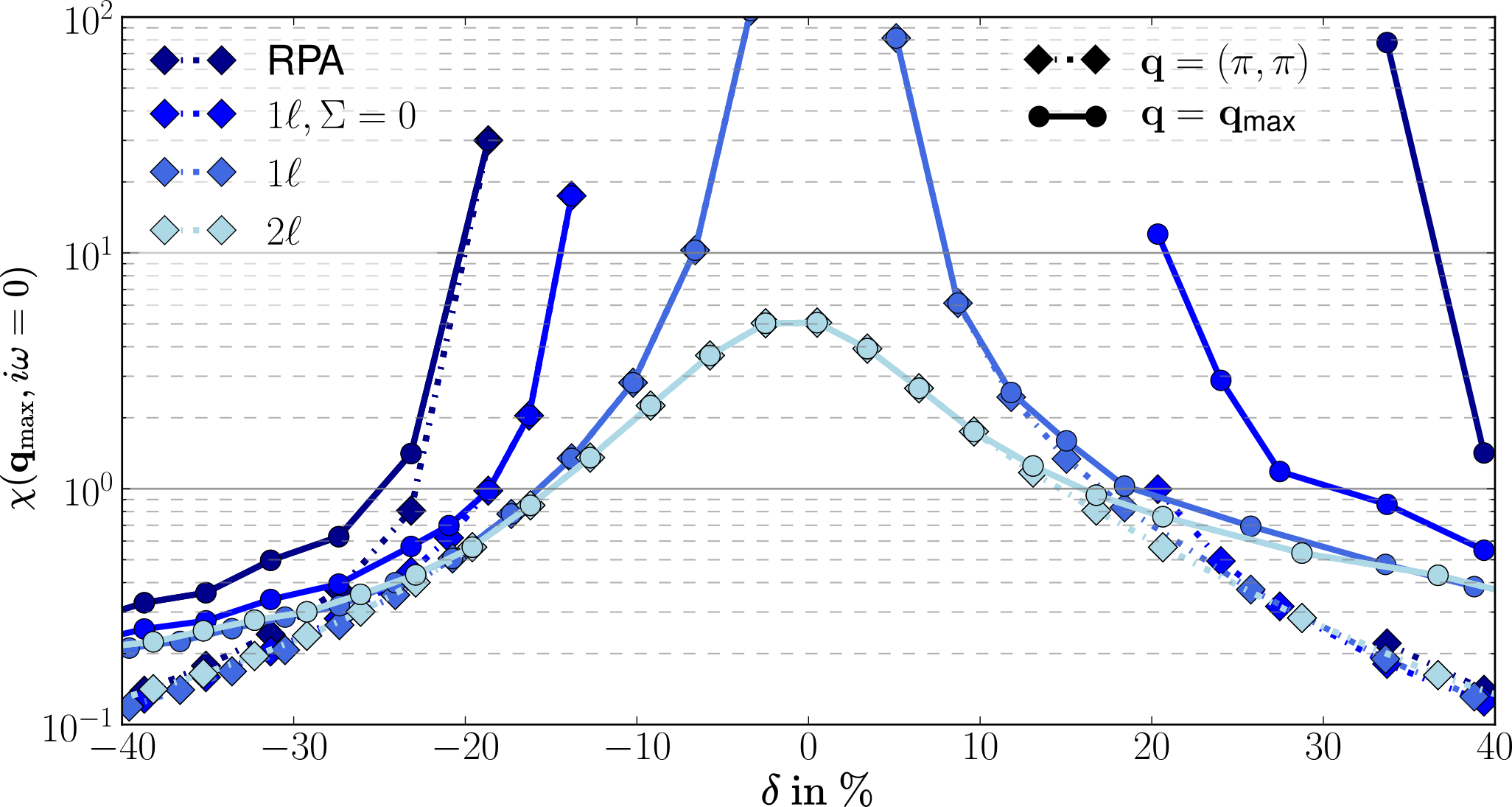}
    \caption{Magnetic $\chi^{\M}(\bf{q}_{\text{max}},i\omega=0)$ and $d$-wave superconducting $\chi^{\SC}(\mathbf{q}=(0,0),i\omega=0)$ susceptibilities as obtained from different approximations: the RPA and fRG in the $1\ell$ truncation (neglecting the 
    self-energy flow),
    for the same parameters as in Fig.~\ref{fig:phasediagram_overview} but for $\beta=10$ since for lower temperatures both RPA and $1\ell$ fRG without self-energy flow do not converge. 
    For comparison, also data with self-energy are shown, for $1\ell$, $2\ell$. 
    }
    \label{fig:RPA_VS_noSig}
\end{figure}
 
The large AF fluctuations, 
whose magnitude grows 
both with increasing interaction $U$ 
and deacreasing temperatures,
originate mostly from the  
ladder diagrams. 
This can be deduced from the comparison to the RPA, 
reproduced by the $1\ell$ fRG flow in a single channel\footnote{If the self-energy is included, the Katanin scheme \cite{Katanin2004} has to be applied.}, see \cfg{RPA_VS_noSig}. 
The susceptibility obtained from the crossed particle-hole ladder of the RPA is drastically lowered by the inclusion of the inter-channel feedback in the 
fRG. This trend is enhanced with increasing 
loop order and can be understood from a diagrammatic point of view: the RPA 
includes only ladder diagrams yielding a magnetic susceptibility of the form  $\chi^{\M,0}/(1-U\chi^{\M,0})$, prone to divergences. 
The $1\ell$ fRG, while still biased towards ladder diagrams, also partially accounts for the screening effects of 
parquet diagrams. 
With increasing loop order, this bias is gradually lifted\footnote{We note that the convergence towards the parquet approximation 
exhibits a characteristic 
oscillatory behavior in the loop order \cite{Tagliavini2019,Chalupa2022}.} as 
diagrams of increasing order in $U$ are 
included. For $\ell \ge 2$, the diagrams of the parquet approximation are accounted for correctly 
up to $O(U^3)$. 
An important role is also played by the imaginary part of the self-energy  
in renormalizing 
the bubble contribution $\chi^{\M,0}$, in particular 
around half filling.
In its evaluation with the Schwinger-Dyson equation, the self-energy itself appears to be controlled by AF fluctuations  
\cite{Hille2020b,Braun22}.

\subsection{Superconducting susceptibility}
\label{sec:superconductivity}

\begin{figure}[h!]

\hfill $s$-wave \hspace{6cm} $d$-wave \hfill
\includegraphics[width=\linewidth]{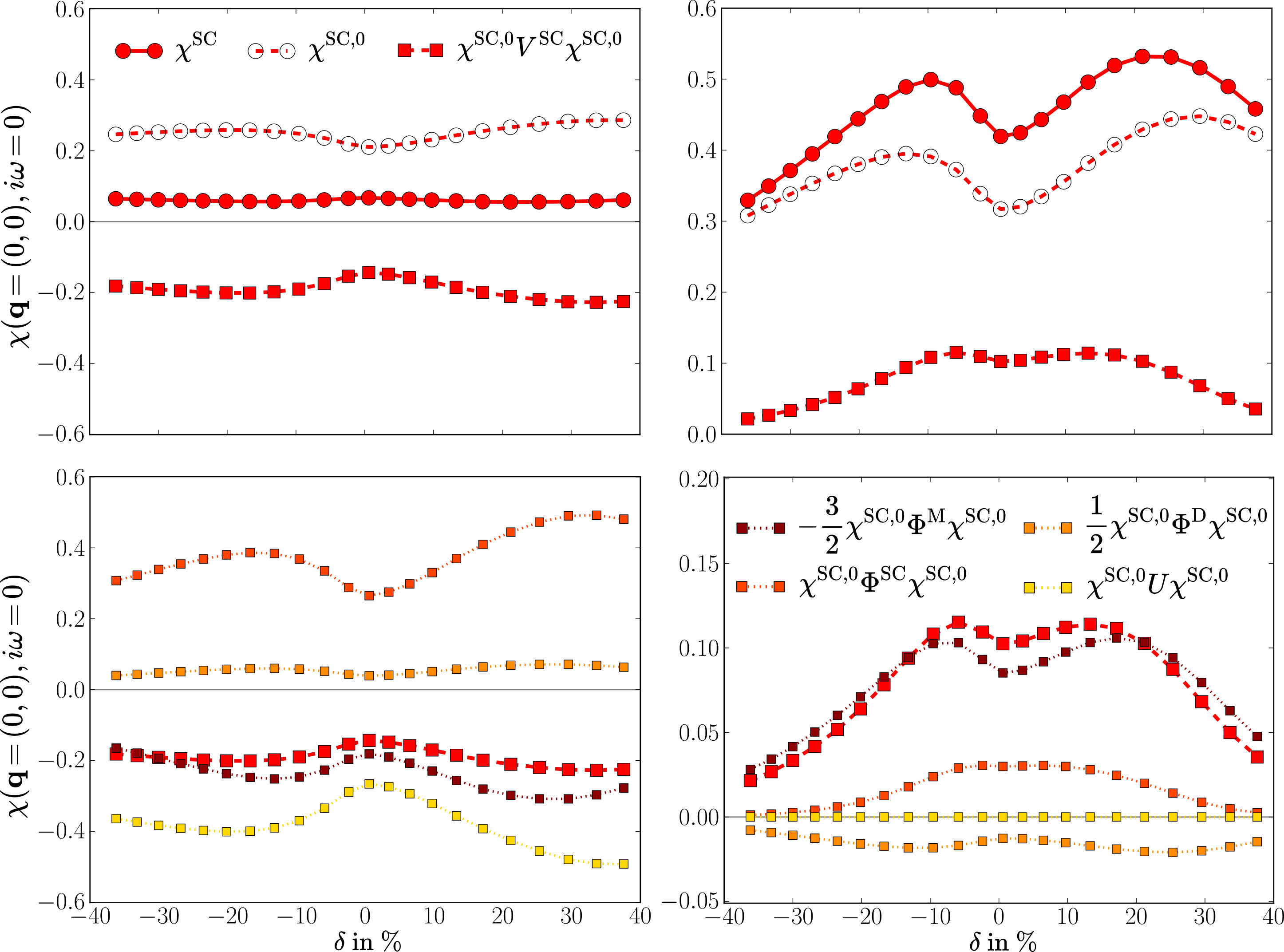}
\caption{Fluctuation diagnostics of the superconducting susceptibility $\chi^{\SC}(\mathbf{q}=(0,0),i\omega=0)$  
of Fig.~\ref{fig:phasediagram_overview}, both for the  $s$- and $d$-wave components (left and right panels respectively).   
Note that although the absolute values of the 
different $d$-wave contributions  
are smaller than 
the respective $s$-wave ones, they add up instead. In contrast, the negative sign of several $s$-wave contributions leads to a partial cancellation. 
}
\label{fig:fullpp_B15_SC}
\end{figure}

In analogy to the analysis performed for 
the magnetic channel, we determine the 
bubble and vertex contributions to the $s$- and $d$-wave superconducting susceptibility $\chi^{\SC}$. 
Inserting the parquet decomposition of $\mathbf{V}^\SC$ into the post-processing Eq. \eqref{eq:susc_postproc}, we obtain
\begin{align}
    \chi^\SC (\mathbf{q},i\omega)
    =& 
    \chi^{\SC,0} (\mathbf{q},i\omega)+ \left[\chi^{\SC,0}\mathbf{\Lambda}^{\mathrm{2PI}}\chi^{\SC,0} \right](\mathbf{q},i\omega)\nonumber \\
    &-\left[\chi^{\SC,0}P^{\xph \rightarrow pp} \mathbf{\Phi}^\M \chi^{\SC,0}\right](\mathbf{q},i\omega) - \frac{1}{2}\left[\chi^{\SC,0} P^{ph \rightarrow pp} \left(\mathbf{\Phi}^\M-\mathbf{\Phi}^\D\right) \chi^{\SC,0}\right] (\mathbf{q},i\omega)\nonumber \\
    &
    + \left[\chi^{\SC,0}\mathbf{\Phi}^{\SC}\chi^{\SC,0}\right](\mathbf{q},i\omega) \;.
    \label{eq:suscsplit_sc}
\end{align}
The different contributions both for $s$- and $d$-wave superconducting susceptibilities, 
evaluated at the maximum of $\chi^\SC$ at $\mathbf{q}_\text{max}=(0,0)$ and $i\omega=0$ , are displayed in Fig.~\ref{fig:fullpp_B15_SC} as a function of the 
doping, for the reference parameters of Fig.~\ref{fig:phasediagram_overview}.
The negative contribution of the $s$-wave $pp$-vertex correction visible in Fig. \ref{fig:fullpp_B15_SC} is expected for a repulsive local interaction. 
Comparing the overall values, the 
$d$-wave $\chi^\SC$ clearly dominates over the $s$-wave one. This is due to the cancellation between the $s$-wave bubble and the vertex contribution 
leading to small overall values with a weak doping dependence, in contrast to the behavior observed in the $d$-wave case where they add up. 
The 
different contributions to the latter 
will be discussed in the following, through a detailed analysis of the terms on the right hand side of Eq.~\eqref{eq:suscsplit_sc}. 

\begin{figure}
    \centering
    \includegraphics[width=0.9\linewidth]{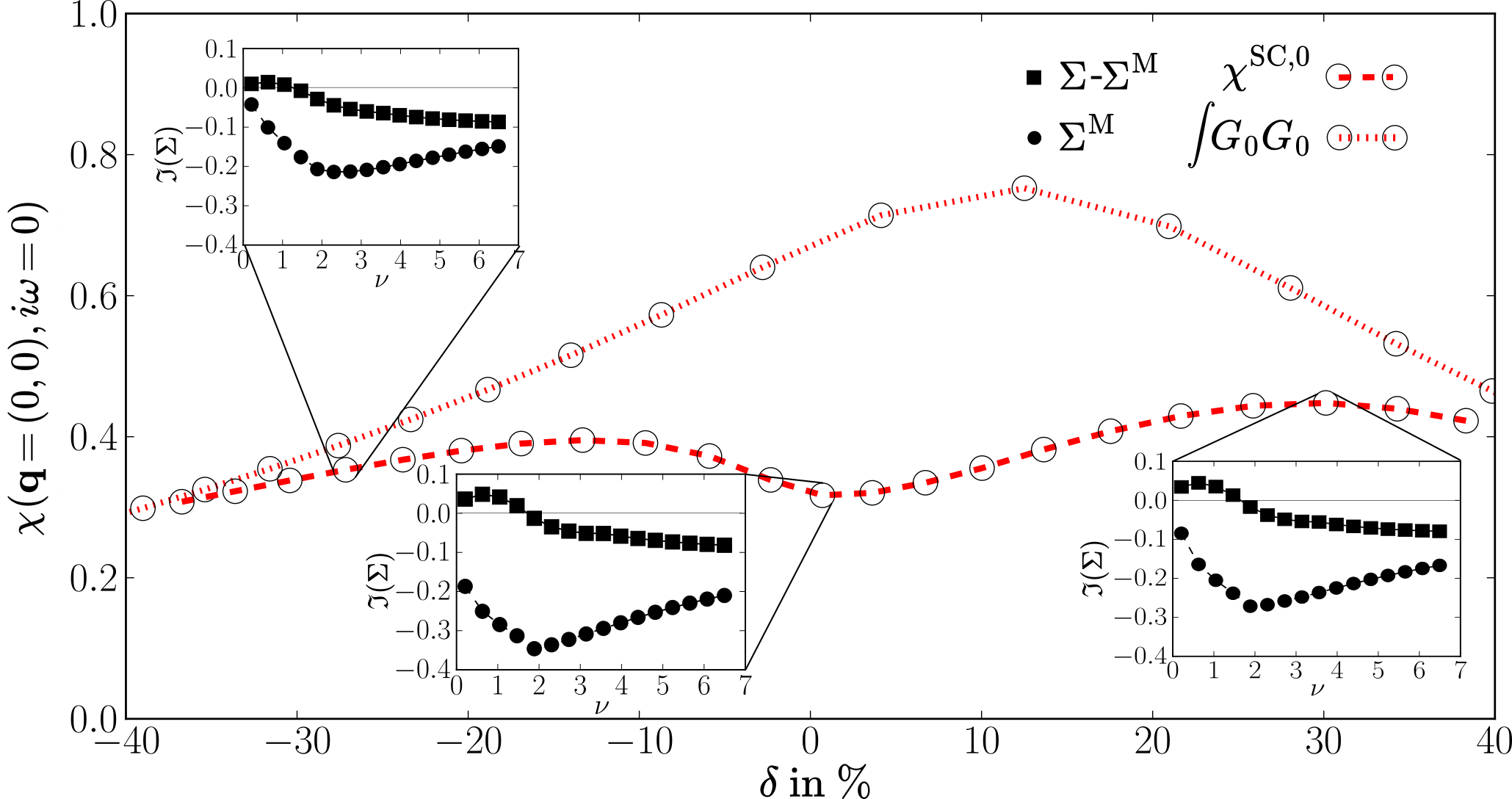}
    \caption{Bubble contribution $\chi^{\SC,0}(\mathbf{q}=(0,0),i\omega=0)$ to the $d$-wave superconducting susceptibility shown in Fig.~\ref{fig:fullpp_B15_SC}, where the results obtained by using the bare Green's function (dotted line) are compared to the ones of the full Green's function (dashed line) including 
    the self-energy.
    The insets show the imaginary part of $\Sigma^\M(\mathbf{k}=(\pi,\pi),i\nu)$ and the sum of all other contributions to the self-energy $\Sigma(\mathbf{k}=(\pi,\pi),i\nu)$, for three representative dopings.
}
    \label{fig:bubbles}
\end{figure}

\subsubsection{$d$-wave bubble contribution to $\chi^{\SC}$}

For the fluctuation diagnostics of the superconducting susceptibility $\chi^\SC$, we start by discussing the results for the bubble contribution. 
For the considered parameters 
($U=3$, $t'=-0.15$, and $\beta=15$), the bare bubble is the dominating contribution in the whole doping range. 
However, when the temperature is lowered, the growth of the bubble is much slower than 
the vertex part (see \cfg{invchi_SC_T_contr}). For large absolute values of the doping $|\delta|$, the corresponding chemical potential $|\mu|$ leads to a decrease of the Green's functions, reflecting the transition to an effectively "uncorrelated" region of the phase diagram.

The origin of the suppression around half filling can be traced back to 
the renormalization of the self-energy and in particular to an enhanced 
imaginary part.  
Without self-energy feedback, the bubble has a single broad maximum around $\delta=10\%$, as shown in \cfg{bubbles}.
Performing a fluctuation diagnostics in analogy to the susceptibilities, the imaginary part can be split \cite{Gunnarsson2016,Schaefer2021b} into the individual contributions according to the parquet equation. Their analysis reveals that all but 
the magnetic contribution remain almost 
constant in doping, while the one from $\Phi^\M_{00}$ increases close to 
half filling,  controlling the imaginary part\footnote{We note that the
slight kink between the 5th and 6th frequency corresponds to the crossing of the low-frequency tensor range and the high-frequency asymptotics of the two-particle vertex.}. 
Similarly to 
the magnetic susceptibility, 
the contribution of $\Phi^\M_{00}$ from the ${\xph}$-channel 
drives the physical behavior 
and is responsible for the double-dome structure induced by the self-energy.

\subsubsection{Vertex corrections to $\chi^{\SC}$} 

In \cfg{invchi_SC_T_contr} we display the temperature dependence of the different contribution to the $d$-wave superconducting susceptibility for $\delta=21.5\%$ (in correspondence of the maximum, see \cfg{fullpp_B15_SC}).
We observe that for temperatures lower than $\beta=15$  
considered above, the vertex corrections given by 
$\chi^{\SC,0}\mathbf{V}^{\SC}\chi^{\SC,0}$ $ (= \chi^\SC - \chi^{\SC,0}$) become more important, whereas the bubble 
stagnates. 
Since the occurrence of a thermodynamic divergence at finite $T$ 
has to be ultimately driven by the vertex and not by the bubble, we now examine the individual contributions 
in more detail.
The doping dependence of the vertex corrections resemble the double dome structure of the bubble. This is not surprising, at least at weak coupling, since they 
involve a matrix-like product in form factors with a convolution in frequency space of two bare bubbles and the vertex itself, see Eq.~\eqref{eq:suscsplit_sc}. 
The largest contribution 
stems from $\mathbf{\Phi}^\M$, followed by $\mathbf{\Phi}^{\SC}$, and finally $\mathbf{\Phi}^\D$. Both contributions from $\mathbf{\Phi}^\M$ and 
from $\mathbf{\Phi}^\SC$
increase when lowering the temperature, with the strongest enhancement for $\mathbf{\Phi}^\SC$. 
In comparison, the contributions from $\mathbf{\Phi}^\D$ and $\Lambda^{2\text{PI}}$ are almost negligible. 
For $\mathbf{\Phi}^\D$, the $d$-wave component is strongly suppressed due to the projection from the $ph$ to the $pp$ channel, while the $s$-wave component is much smaller.
$\Lambda^{2\text{PI}}_{mn}$ is instead suppressed by the vanishing mixed $s$- and $d$-wave bubbles at $\mathbf{q}=0$.
    
\begin{figure}
    \centering
        \includegraphics[width=.9\linewidth]{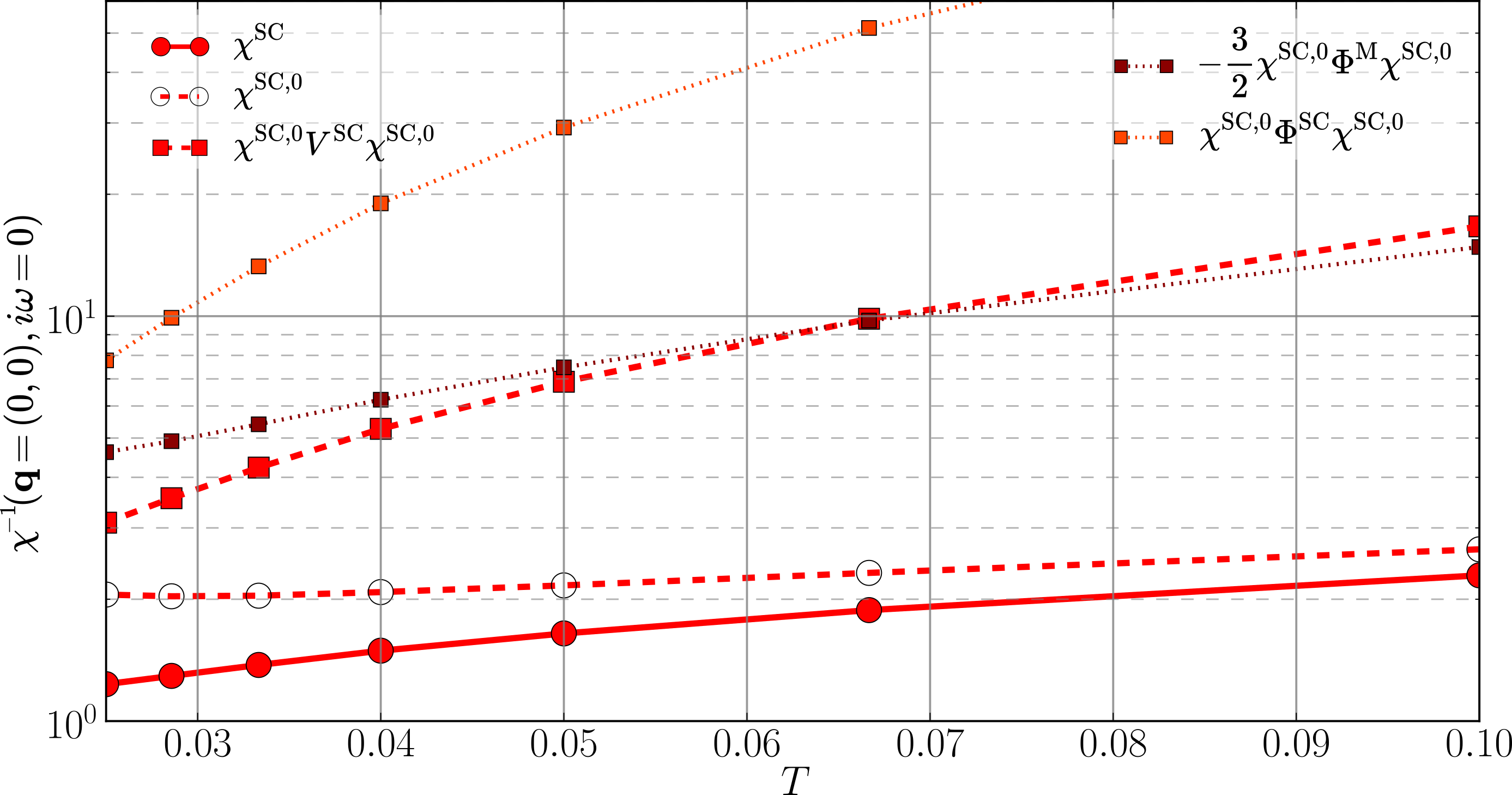}
        \caption{Temperature dependence 
        of the contributions to the $d$-wave superconducting susceptibility $\chi^{\SC}(\mathbf{q}=(0,0),i\omega=0)$ of Fig.~\ref{fig:fullpp_B15_SC}, for $\delta=21.5\%$ corresponding to the maximal value. 
        }
        \label{fig:invchi_SC_T_contr}
\end{figure}

In order to investigate the influence of the leading $\mathbf{\Phi}^\M$ on the $d$-wave superconducting susceptibility, we 
analytically estimate its contribution (the details of the derivation can be found in Appendix \ref{sec:analyt})
\begin{align}
    &\frac{3}{2}\left[\chi^{\SC,0}\mathbf{\Phi}^\M\chi^{\SC,0}\right]_{11}(\mathbf{q},0)  
    \approx \sum\limits_{\substack{i\nu,i\nu'}}\Pi^{\SC}_{11}(\mathbf{q},0,i\nu) \Pi^{\SC}_{11}(\mathbf{q},0,i\nu') \nonumber \\
    &\quad\cdot\int_{\text{BZ}} \text{d}\mathbf{p} \left(-\frac{1}{2}\left(\cos{p_x}+\cos{p_y}\right)-\left(\cos{(p_x-q_x)}+\cos{(p_y-q_y)}\right)\right) \nonumber \\
    &\qquad\qquad\quad\;\;\cdot\Phi^\M_{00}(-(i\nu+i\nu'),i(\nu-i\nu')/2,(i\nu'-i\nu)/2,\mathbf{p}), \phantom{\int_{\text{BZ}}}
    \label{eq:magnetic_contr_to_chi_sc}
\end{align}
where the components are indicated explicitly. 
The weighting factor in the bracket
accounts for the form-factor projections $P^{ph\rightarrow pp}$ and $P^{\xph\rightarrow pp}$  between the channels.
The bubble $\Pi^{\SC}_{11}$ is maximal at $\mathbf{q}=(0,0)$, independently of the doping. 
$\Phi^\M_{00}(0,i\nu,i\nu',\mathbf{p})$ has its maximum at $\mathbf{p}=(\pi,\pi)$ around half filling, in correspondence of dominant commensurate AF fluctuations. 
At small dopings, 
the maximum of the weighting factor coincides with the maximum of the bubble at $\mathbf{q}=(0,0)$ and the maximum of $\Phi^\M_{00}$ at $\mathbf{p}=(\pi,\pi)$. 
This is no longer the case for larger dopings, where 
the maximum of $\Phi^\M_{00}$ 
occurs at the incommensurate wave vector $\mathbf{p}=(\pi-\delta,\pi)$. 
The weighting factor 
no longer favours the maximum of $\Phi^\M_{00}$, which is already suppressed with respect to its value at half filling (note that for 
$\mathbf{q}=(0,0)$, the contributions from $P^{ph\rightarrow pp}$ and $P^{\xph\rightarrow pp}$ are 
equal, justifying their combined treatment in the fluctuation diagnostics).
This overall behavior as a function of the doping 
is modified by the presence of the bubble in Eq. \eqref{eq:magnetic_contr_to_chi_sc}, leading to a double dome structure 
qualitatively similar to one of the bubble itself.
In particular, around half filling, where the AF fluctuations are commensurate and strongest in size, the bubble is suppressed. 
The reduction is however less pronounced, due to the prominent maximum of $\Phi^\M_{00}$. At the same time, the decrease at larger dopings is stronger as a result of the weaker  and less favoured incommensurate fluctuations.

The next to leading 
contribution to the vertex correction of $\chi^{\SC}$ is the one from $\mathbf{\Phi}^{\SC}$ itself (see Fig. \ref{fig:invchi_SC_T_contr}). 
The $s$-wave component of $\mathbf{\Phi}^{\SC}$, which includes the $pp$-ladder diagrams, is much larger than the $d$-wave one for all dopings. However, its contribution vanishes at $\mathbf{q}_{\text{max}}=(0,0)$ due to the mixed $s$- and $d$-wave bubble in $\chi^{\SC,0}\mathbf{\Phi}^{\SC}\chi^{\SC,0}$.
A finite $d$-wave component is generated only by diagrams where neither the two incoming nor the two outgoing legs 
combine into a purely local bare vertex, i.e. the ones included in the inner frequency box \cite{Fraboulet2022}. 
The lowest order diagram 
of $\mathcal{O} (U^4)$ is fully included only in the $2\ell$ approximation, since in the $1\ell$ flow only the Green's functions of  
the $pp$-bubble are replaced by their derivative, see \cfg{xph_in_pp}.

\begin{figure}
    \centering
    \includegraphics[width=0.6\linewidth]{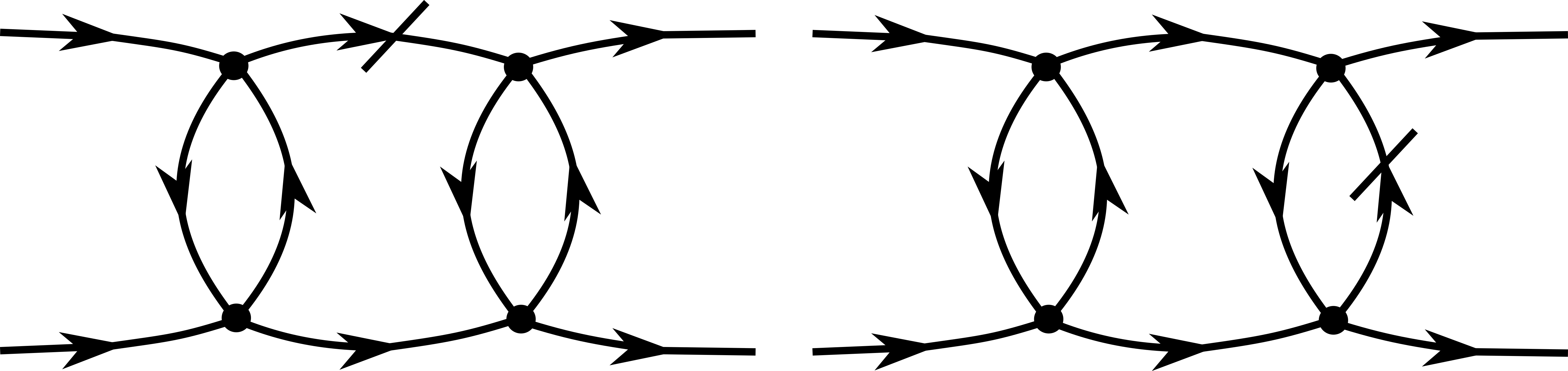}
    \caption{The diagram on the left is included in the $1\ell$ approximation of the flow, 
    whereas the one on the right is introduced 
    only in 
    $2\ell$. } 
    \label{fig:xph_in_pp}
\end{figure}

\subsubsection{Multiloop effects}
\label{sec:multiloop}

We proceed by investigating the effect of higher loop orders on the different contributions to the $d$-wave $\chi^{\SC}$, the results are reported in  \cfg{SCcontr_loop}.
In the analysis of the magnetic susceptibility, 
we saw that 
multiloop corrections 
(as well as the self-energy feedback) quench the effect of the ladder diagrams in the $\xph$-channel, suppressing the tendency towards a magnetic instability.
For the $d$-wave superconducting $\chi^{\SC}$, we  find the opposite behavior: 
removing the bias towards ladder diagrams in the multiloop scheme leads to an 
\textit{increase} of 
$\mathbf{\Phi}^{\SC}$. 
All contributions 
present a distinct enhancement from the $1\ell$ (also with Katanin substitution) to the $2\ell$ results, followed by a saturation for higher loop orders. This behavior 
confirms the $2\ell$ approximation to be  
an optimal compromise between quantitative accuracy at loop convergence and reduced numerical effort, besides 
being correct up to $\mathcal{O} (U^3)$.

By far the largest relative increase occurs in 
the $\mathbf{\Phi}^{\SC}$ component which more than doubles between the Katanin approximation and the $2\ell$ result (see also discussion above). 
Though its share in the sum 
is modest for the considered parameters, in particular when compared to its counterpart in the magnetic channel, it increases rapidly with $\beta$ as shown in \cfg{invchi_SC_T_contr}. 
(their analytic expressions for the lowest order diagrams 
$\mathbf{q}=(0,0)$
are derived 
in Appendix \ref{sec:analyt}). 

\begin{figure}
    \centering
        \includegraphics[width=.9\linewidth]{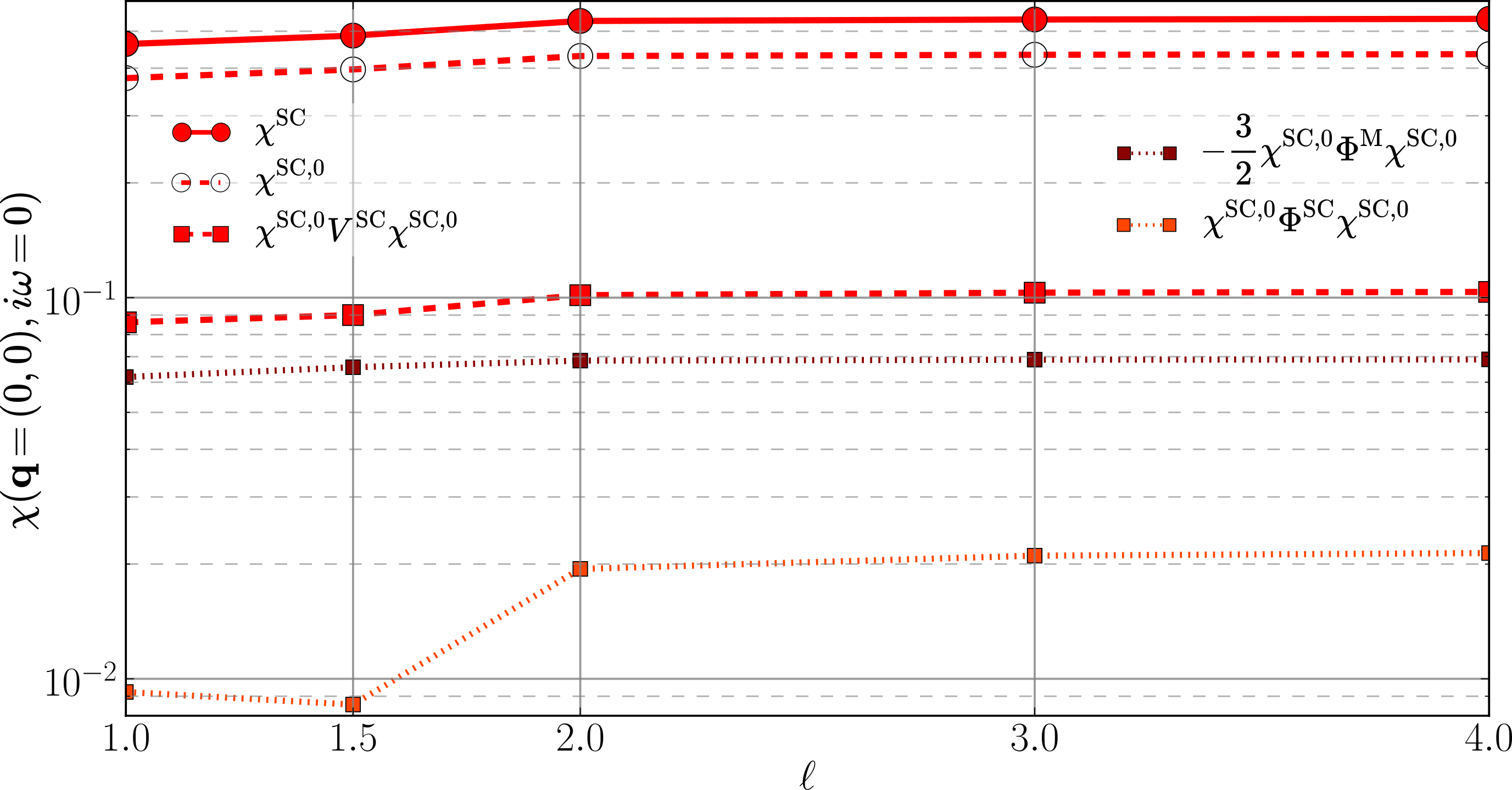}
        \caption{Multiloop dependence of the contributions  to the $d$-wave superconducting  susceptibility $\chi^{\SC}(\mathbf{q}=(0,0),i\omega=0)$ of Fig.~\ref{fig:fullpp_B15_SC}, for $\delta=21.5$.  
        For completeness also $1\ell$ and $1.5 \ell$ for the Katanin substitution are shown.}
        \label{fig:SCcontr_loop}
\end{figure}

\subsubsection{Fluctuation diagnostics in terms of bosonic fluctuations}

\begin{figure}
     \centering
     \includegraphics[width=0.9\linewidth]{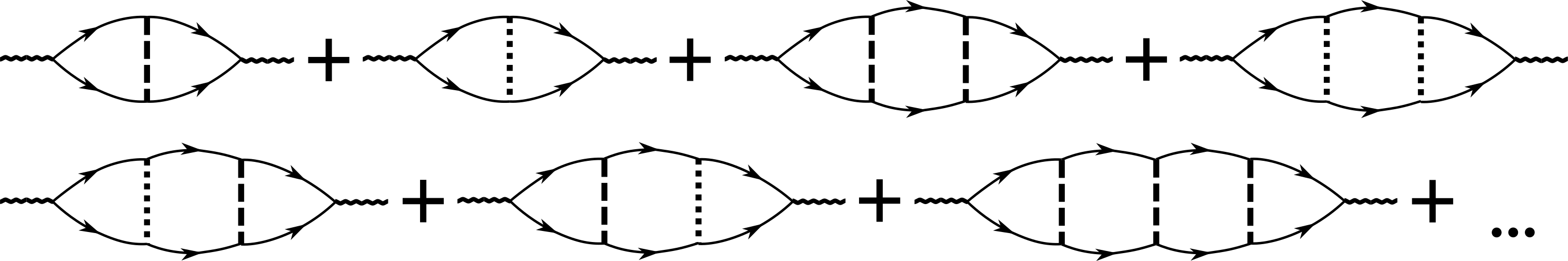}
     \caption{Bosonic fluctuation contributions to the $d$-wave superconducting susceptibility, where the dashed and dotted lines represent $s$-wave magnetic and density bosonic propagators  
     respectively. 
     }
     \label{fig:fluct_diag_boson}
 \end{figure}

\begin{figure}
     \centering
     \includegraphics[width=\linewidth]{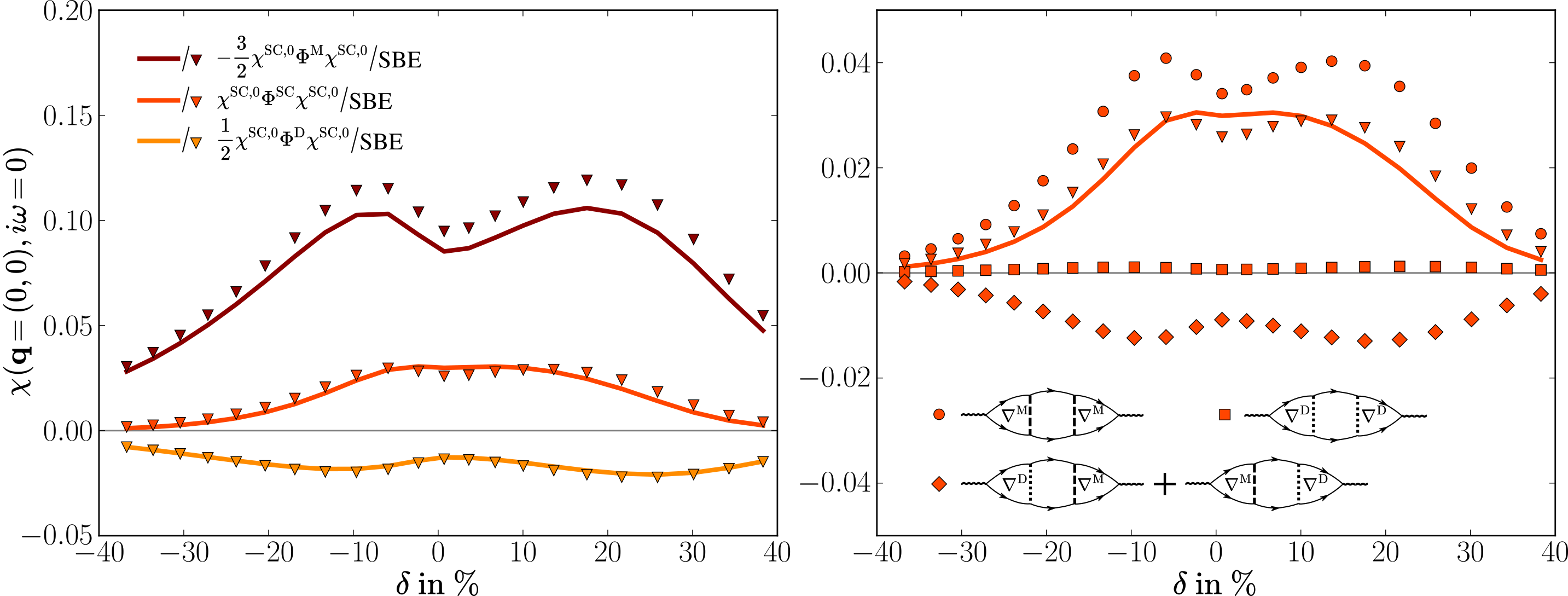}
     \caption{Fluctuation diagnostics based on the exchange of bosonic fluctuations (see text for the details), for the vertex corrections of the $d$-wave superconducting susceptibility of Fig.~\ref{fig:phasediagram_overview}, with a comparison to the one obtained from the parquet decomposition.
     }
     \label{fig:fluct_diag_channel_vs_boson}
 \end{figure}

In addition to 
the fluctuation diagnostics based on the parquet decomposition, we perform also a decomposition 
based on the exchange of bosonic fluctuations \cite{Krien2019,Krien2021}. 
Using 
the single-boson-exchange (SBE) formalism \cite{Krien2019}, we determine the $d$-wave superconducting susceptibility as a series of boson exchanges $\nabla^{\M/\D}$ \cite{Bonetti2021}, replacing the post-processing relation \eqref{eq:susc_postproc}
\begin{align}
    \left[\chi^{\SC,0}V^\SC\chi^{\SC,0}\right]_{11}(\mathbf{q},0)  \approx &-\frac{3}{2}\left[\chi^{\SC,0} \nabla^{\M} \chi^{\SC,0}\right]_{11} (\mathbf{q},0) +\frac{1}{2}\left[\chi^{\SC,0} \nabla^{\D} \chi^{\SC,0}\right]_{11}(\mathbf{q},0)  \nonumber \\
    &+\left(-\frac{3}{2}\right)^2 \left[\chi^{\SC,0} \nabla^{\M} \chi^{\SC,0} \nabla^{\M} \chi^{\SC,0}\right]_{11} (\mathbf{q},0) \nonumber \\
    &+ \left(\frac{1}{2}\right)^2\left[\chi^{\SC,0} \nabla^{\D}\chi^{\SC,0} \nabla^{\D}\chi^{\SC,0}\right]_{11} (\mathbf{q},0) \nonumber \\
    &-2\frac{3}{2}\frac{1}{2}\left[\chi^{\SC,0} \nabla^{\D}\chi^{\SC,0} \nabla^{\M}\chi^{\SC,0}\right]_{11} (\mathbf{q},0) \nonumber \\  
    &+\left(-\frac{3}{2}\right)^3 \left[\chi^{\SC,0} \nabla^{\M} \chi^{\SC,0} \nabla^{\M} \chi^{\SC,0} \nabla^{\M} \chi^{\SC,0}\right]_{11}(\mathbf{q},0)  + \dots
\end{align}
where $\left[\chi^{\SC,0} \nabla^{\M/\D} \chi^{\SC,0}\right]_{11}(\mathbf{q},0)\approx \left[\chi^{\SC,0}\mathbf{\Phi}^{\M/\D}\chi^{\SC,0}\right]_{11}(\mathbf{q},0)$ for the single magnetic / density boson exchange contributions in the first line (the difference to their parquet counterparts lies in the SBE rest function), 
all multiboson exchange processes are contained in the contribution to $\mathbf{\Phi}^\SC$. This alternative decomposition is illustrated diagrammatically  
in Fig.~\ref{fig:fluct_diag_boson}, the results for the corresponding fluctuation diagnostics together with a comparison to the previous data are displayed in Fig.~\ref{fig:fluct_diag_channel_vs_boson}.
As we do not have direct access to $\nabla^{\M/\D}$ within the parquet decomposition, we construct them from the high-frequency asymptotics \cite{Fraboulet2022,HeinzelmannThesis}. 
The contribution of the magnetic channel 
encoded in 
$\nabla^{\M}$ 
is in qualitative agreement with the contribution of $\mathbf{\Phi}^{\M}$, the quantitative differences are due to the effect of the SBE rest function which is most pronounced in the leading channel \cite{Bonetti2020,Fraboulet2022}. 
In fact, this does not affect the contribution of $\nabla^{\D}$ that agrees very well with the one of $\mathbf{\Phi}^{\D}$ for all dopings.
For the 
contributions to  
$\mathbf{\Phi}^{\SC}$, we focus on the two-boson processes for simplicity, since higher order multiboson diagrams are not fully included at the $2\ell$ level. 
In Fig. \ref{fig:fluct_diag_channel_vs_boson}, the results for the sum of the two-boson exchange processes reproduce the contribution of $\mathbf{\Phi}^{\SC}$ fairly well.
We finally note that the magnetic and density fluctuations have opposite signs: while the magnetic fluctuations enhance the $d$-wave superconducting susceptibility, 
the density ones reduces it (see also the negative contribution of $\mathbf{\Phi}^\mathrm{D}$ in Fig. \ref{fig:fullpp_B15_SC}), recalling to a given extent, the trend observed for the decomposition of the self-energy in D$\Gamma$A calculations \cite{Gunnarsson2016}. Similarly, the multiboson contributions including an odd number of density bosons  
are negative. 
This alternative decomposition offers hence a physically intuitive interpretation of the underlying fluctuations.

\subsubsection{Dependence on Hubbard parameters $U$ and $t'$}

\begin{figure}
         \centering
         \includegraphics[width=\linewidth]{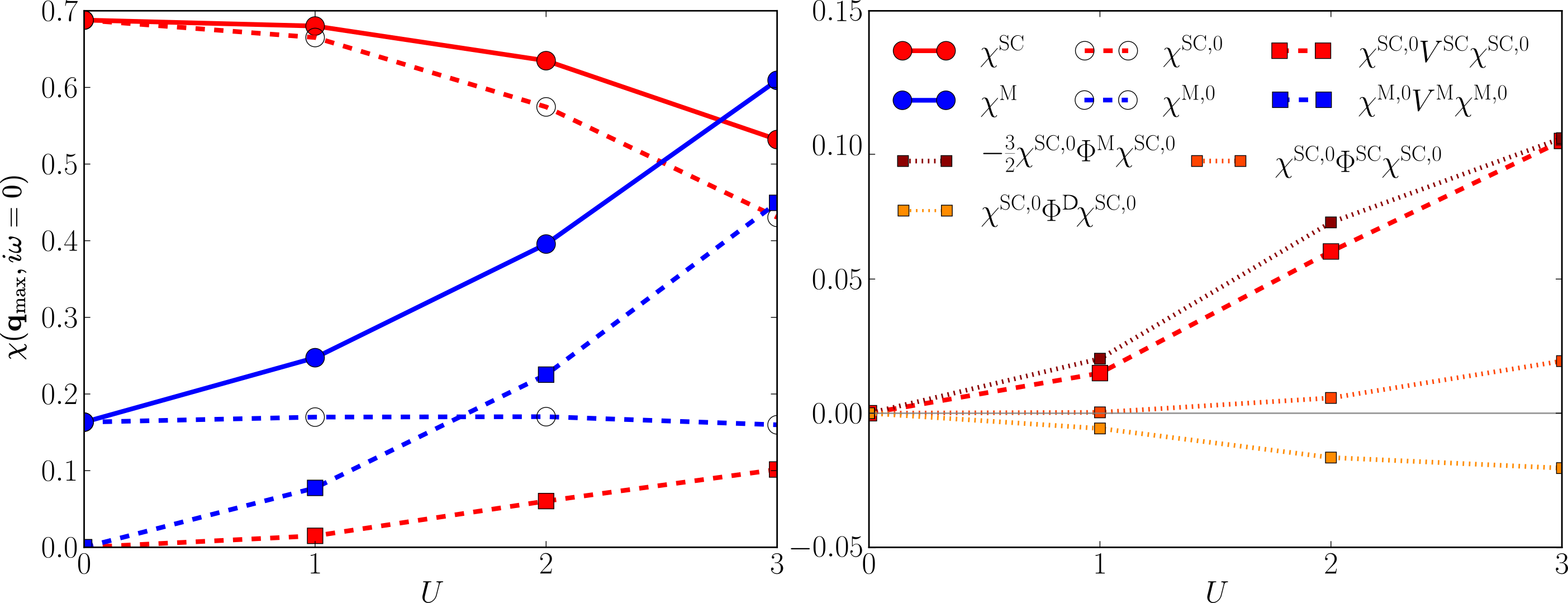}
         \caption{Dependence on the interaction strength $U$ of the different contributions to the $d$-wave superconducting susceptibility $\chi^{\SC}(\mathbf{q}=(0,0),i\omega=0)$ of Fig.~\ref{fig:fullpp_B15_SC}, for $\delta=21.5$.  
         The vertex contributions are shown in the left, the bubble and vertex corrections on the right panel, where the data for the magnetic susceptibility are reported for comparison.
         }
         \label{fig:Us}
\end{figure}

We now 
explore the role of the 
parameters $U$ and $t'$, which have been fixed for the previous analysis. In Fig.~\ref{fig:Us} we present a fluctuation diagnostics for $\delta=21.5\%$, where $\chi^{\SC}$ reaches its maximum for $\beta=15$.
For smaller values of the interaction, a qualitatively different picture with respect to the default $U=3$ emerges: 
$\chi^{\SC}$ is larger but entirely bubble driven, while the vertex corrections are almost negligible due to lack of AF fluctuations. For the same reason, the bubble itself is larger due to the reduced self-energy corrections.
As a consequence, the dome structure is absent at 
weak coupling, at least in the temperature regime accessible to our calculations.
As the interaction strength is increased, $\Phi^\M_{00}$ grows together with the vertex corrections and the self-energy, which in turn dampens the bubble, inducing thus a more correlated regime 
by $U=3$.
The largest relative increase is observed in ${\Phi}^{\SC}$, 
whose lowest order diagram scales with 
$\mathcal{O} (U^4)$. 

\begin{figure}
    \centering
    \includegraphics[width=\linewidth]{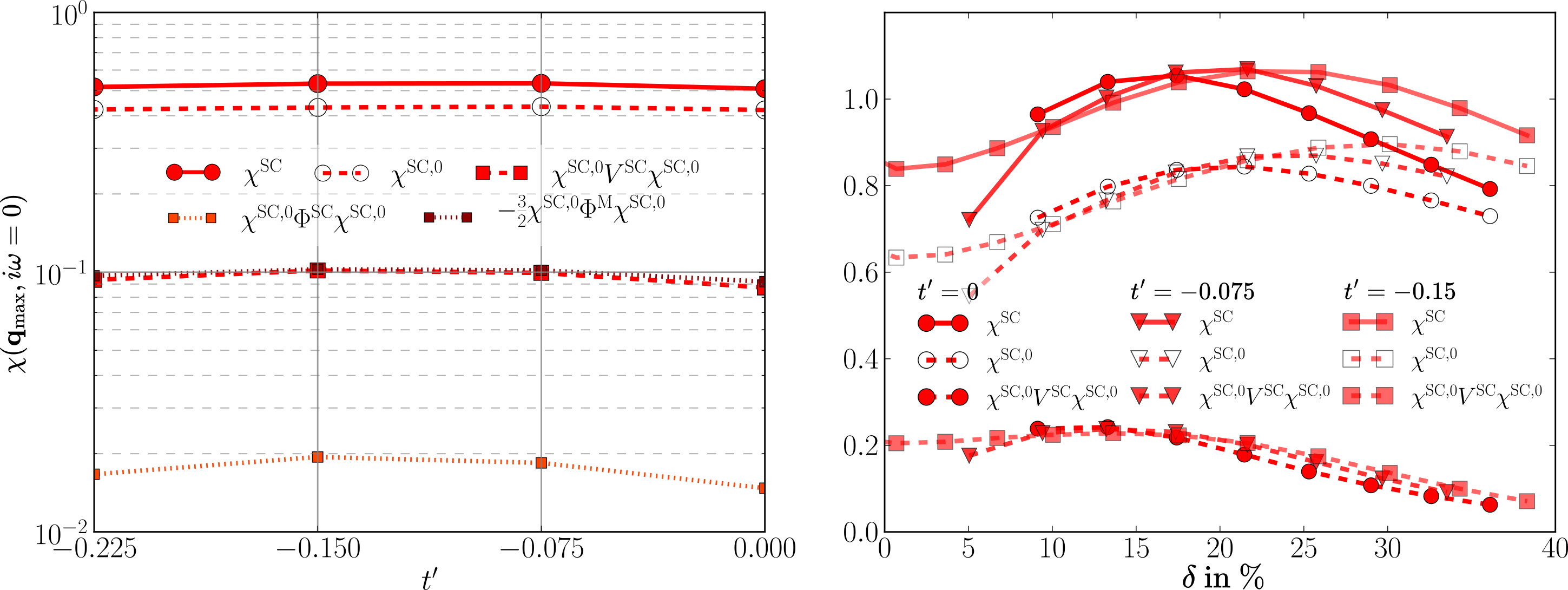}
    \caption{Dependence on the next-nearest neighbor hopping amplitude $t'$ of the different contributions to the $d$-wave superconducting susceptibility $\chi^{\SC}(\mathbf{q}=(0,0),i\omega=0)$ of Fig.~\ref{fig:fullpp_B15_SC}. In the left panel, a fixed value of the doping $\delta=21.5$ considered, while in the right one the behavior with doping is reported. 
    } 
    \label{fig:TPs}
\end{figure}

The left panel of Fig. \ref{fig:TPs} shows the result for different values of $t'$, for the same doping and $U=3$ ($\beta=15$). With increasing next-nearest neighbor hopping amplitudes, the growing asymmetry leads to an enhancement of the dome at hole doping and to a suppression of the one in the electron-doped regime.
However, the overall doping dependence of the different components turns out to be rather weak. 
Concerning the behavior as a function of doping displayed in the right panel of Fig. \ref{fig:TPs}, 
the maxima of both bubble and vertex contribution to the superconducting susceptibility shift to larger values with increasing $|t'|$. This effect is more pronounced for the bubble, whose maximum increases with $|t'|$ whereas the one of the vertex exhibits the largest value for $t'=0$. 
We note that a large $t'$ hopping amplitude 
of $\mathcal{O}(1)$ effectively alters the lattice geometry leading to 
frustration that dampens the AF peak as well as all other susceptibilities. 
The optimal doping is hence determined by the maxima of both contributions to 
be as close as possible in doping and their sum (of the decreasing vertex and increasing bubble) to be maximal. We note that these effects are very small on a quantitative level, as observed in Fig. \ref{fig:TPs}. This also means that the results of our analysis can be considered generically applicable to the weak-coupling regime of the 2D Hubbard model.

\subsubsection{Summary: What drives $d$-wave superconductivity?}

While the divergence of the magnetic channel can easily be traced to 
ladder diagrams and nesting of the Fermi surface, the picture is much more complex for the 
($d$-wave) superconducting channel, characterized by a subtle interplay between various competing effects.
In the following we will summarize the physical picture resulting from the our detailed analysis.

As a first prerequisite for substantial vertex corrections in the superconducting susceptibility of the Hubbard model are strong AF fluctuations \cite{Dong2022,Krien2021}. These are generated for sufficiently strong interactions. In addition, both the doping and the next-nearest neighbor hopping amplitude has to be sufficiently small, since they suppress 
the AF fluctuations 
and shifts its maximum towards incommensurate wave vectors.

We now discuss the interaction effects in more detail, considering the various components contributing to the susceptibility as introduced in the fluctuation diagnostics and the self-energy.
We saw that feedback of its imaginary part to $\mathbf{\Phi}^\M$ leads to a double dome structure with a local minimum 
close to half filling. 
Hence, the large AF fluctuations around half filling drive both the importance of the magnetic channel in the vertex corrections and the screening induced by the self-energy.
Another relevant mechanism is provided by the interplay between the magnetic, density, and superconducting channels and their mutual feedback:
a large $\Phi^\M_{00}$ enhances the other channels, whose feedback on the other hand controls the divergent tendencies of $\xph$-ladders.
With increasing multiloop order, $\Phi^\M_{00}$ is renormalized, while the other $d$-wave channels and in particular the superconducting one, gain in importance 
as they rely on the input from other channels. 

\subsection{Density susceptibility}

\begin{figure}[h!]
\hfill $s$-wave \hspace{6cm} $d$-wave \hfill
\includegraphics[width=\linewidth]{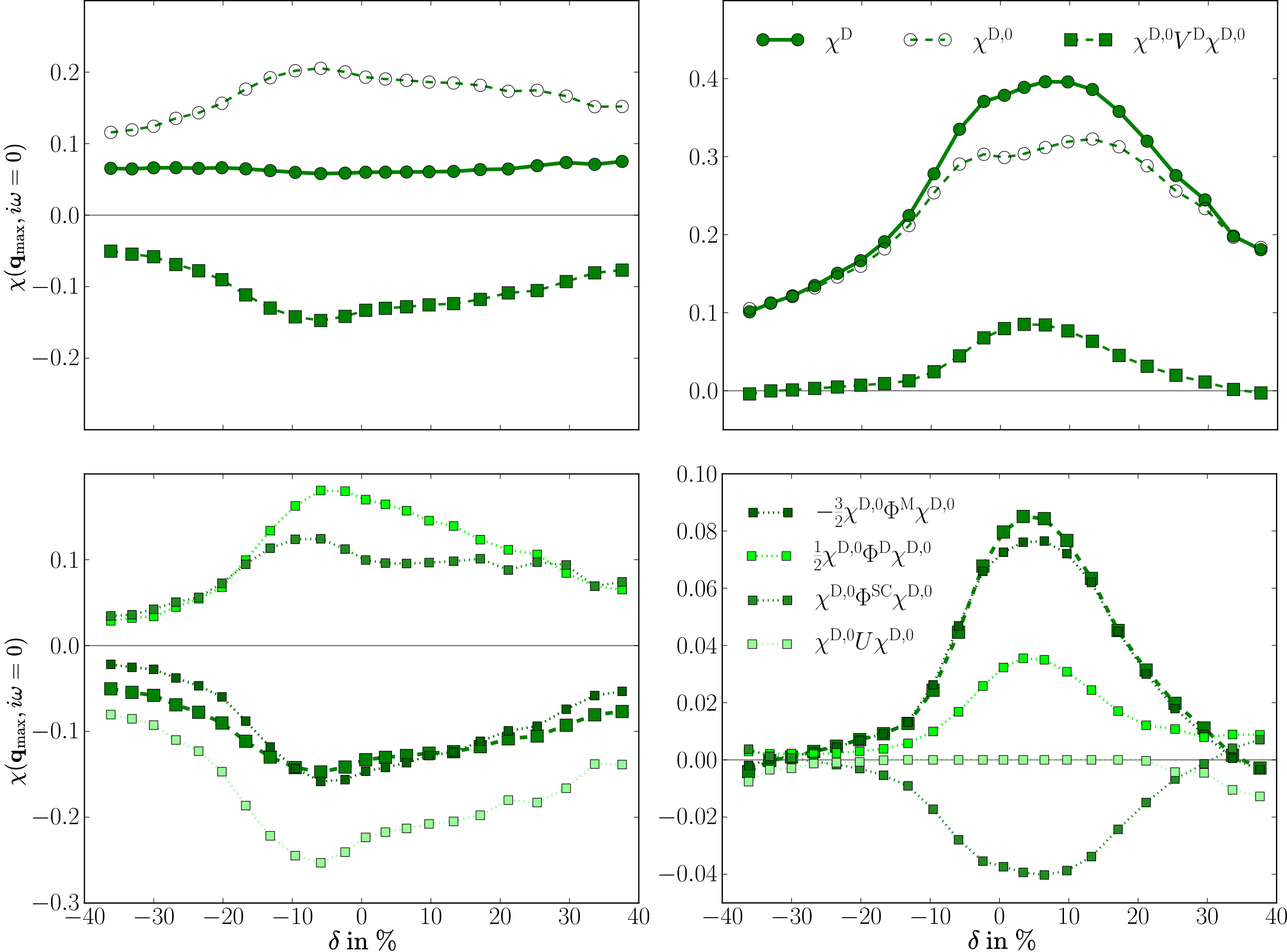}
\caption{Fluctuation diagnostics of the density susceptibility $\chi^{\D}(\mathbf{q}_{\text{max}},i\omega=0)$ of Fig.~\ref{fig:phasediagram_overview}, both for the  $s$- and $d$-wave components (left and right panels respectively). 
Note that although the absolute values of the different $d$-wave contributions are smaller than the respective $s$-wave ones, they add up instead. In contrast, the negative sign of several $s$-wave contributions leads to a partial cancellation. 
}
\label{fig:pp_susc_D}
\end{figure}

\begin{figure}
    \centering
    \includegraphics[width=\linewidth]{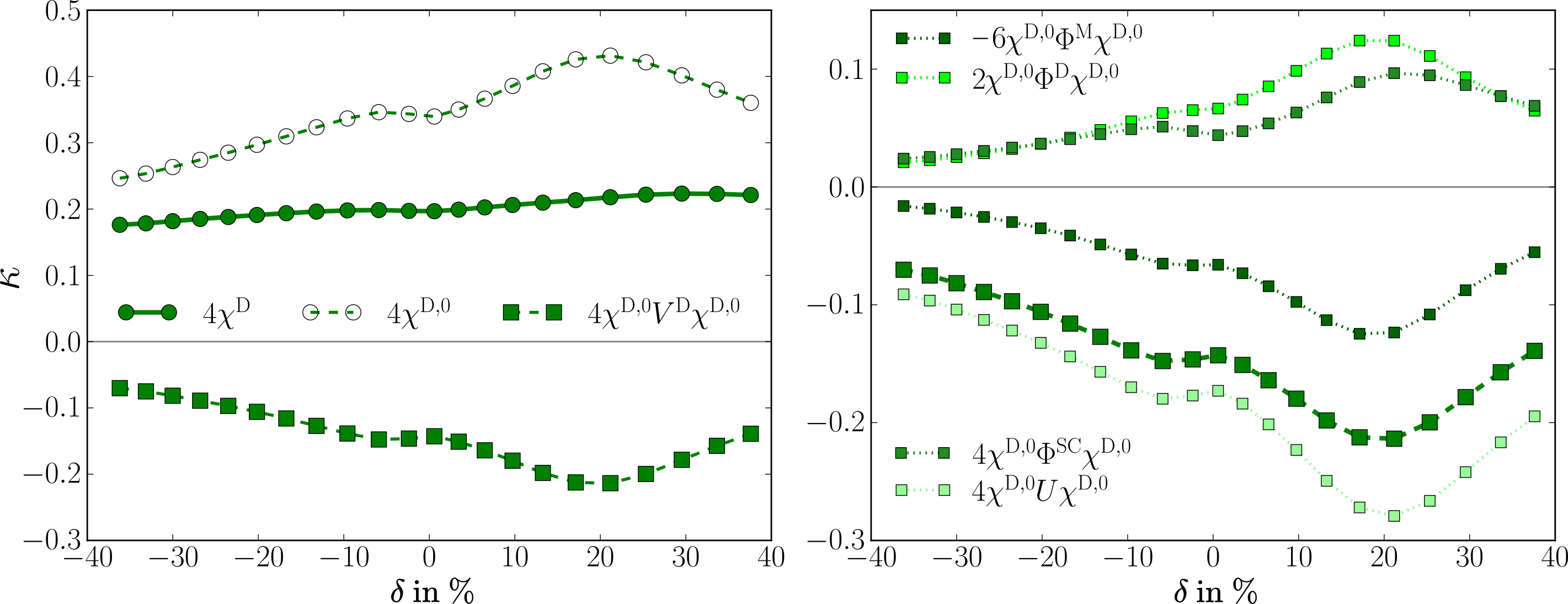}
    \caption{Compressibility $\kappa=4\chi^\D(\mathbf{q}=(0,0),i\omega=0)$ for the same parameters as in Fig.~\ref{fig:phasediagram_overview}.}
    \label{fig:kappa}
\end{figure}

For completeness, we briefly discuss also the results for the diagnostics of the fluctuation for the $d$-wave density susceptibility, 
shown in Fig.~\ref{fig:pp_susc_D} 
for the same parameters as in Fig.~\ref{fig:phasediagram_overview}. For $\chi^{\D}$ significant vertex contributions are observed only around half filling for the $d$-wave component.
In contrast, the $s$-wave components (see Fig.~\ref{fig:pp_susc_D}) of the bubble and vertex corrections are of a similar size but opposing sign.
Their mutual cancellation leads to a very small overall $s$-wave density susceptibility $\chi^{\D}_{00}$ with little dependence on $\delta$, a behavior similar to the superconducting one discussed in the previous section.
We also note that the wave vector corresponding to the maximal value shifts from $(\pi,\pi)$ around half filling to incommensurate values at finite doping (see also Fig.~\ref{fig:postprocM_alongBZ} in Appendix \ref{app:altfig2}). 
Finally, 
we also consider the charge ordering, i.e. the charge density wave susceptibility at $\mathbf{q}=(0,0)$. 
This can be directly related to the ($s$-wave) charge compressibility $\kappa=4\chi^{\D}$. The results are displayed in Fig.~\ref{fig:kappa}.

\section{Conclusions and outlook}
\label{sec:conclusions}

We have computed the magnetic, density, and superconducting pairing susceptibilities of the two-dimensional Hubbard model at weak coupling by using 
the recently developed multiloop extension of the functional fRG.
Analyzing 
the different fluctuation channels and their relative impact on the observed physical behavior, we investigated their evolution with temperature, interaction strength, and loop order both in the electron as well as in the hole doped regime. 
In particular, we determined the impact of vertex corrections on both $s$- and $d$-wave components of the corresponding physical susceptibilities and identified
relevant effects.
Our quantitative fluctuation diagnostics analysis allowed us to trace the influence of AF fluctuations on the $d$-wave superconductivity and provide an analytical understanding for the observed behavior. 

A promising route for future studies is provided by  
the single boson exchange (SBE) formalism 
\cite{Krien2019}, which significantly reduces the numerical effort of treating the high-frequency asymptotics of the flowing vertices.
The application to 
the $1\ell$ fRG flow \cite{Bonetti2021,Fraboulet2022} represents the starting point for multiboson and multiloop extensions of fRG-based algorithms.
In particular, the SBE formulation appears suited for the combination of the multiloop fRG with the DMFT in the so-called DMF$^2$RG approach\cite{Taranto2014}. This would allow to more easily access the non-perturbative regime on a quantitative level even at lower temperatures so far unaccessible.

From a broader perspective, recent advances in different computational methods led to controlled numerical solutions of the Hubbard model in various regions of the phase diagram. Despite the consensus on the general phase diagram of the model, there remain many open challenges such as the determination of the precise location of phase boundaries and in particular the parameter region displaying $d$-wave superconductivity. The physics of models with more orbitals and/or longer ranged interactions remains largely unexplored with accurate methods, and reliable algorithms are yet to be developed. 
Recent years have seen a rapid progress in this direction \cite{Qin2021,Schaefer2021a}. Nevertheless, the development of new, generally applicable numerical techniques which overcome the shortcomings of existing methods 
remains the most urgent need for addressing the open issues. 

\section*{Acknowledgements}
The authors thank P. M.~Bonetti, P.~Chalupa-Gantner, C.~Hille, C.~Honerkamp, A.~Kauch, 
G.~Rohringer, T.~Sch\"afer, and D.~Vilardi for valuable discussions. 

\paragraph{Funding information}
We acknowledges financial support from the Deutsche Forschungsgemeinschaft (DFG) through Project No. AN 815/6-1 (S.H. and S.A.) and
from the Austrian Science Fund (FWF) through the Project I 5868 (part of the FOR 5249 [QUAST] of the DFG, A.T.).

\begin{appendix}

\section{Vertices in channel specific notation}
\label{app:notation}

Each class of diagrams reducible in a specific diagrammatic channel $r$ lends itself to a different parametrization based on the bosonic transfer momentum. When quantities of different channels are added as in the two-particle vertex, they must either be expressed in purely fermionic momenta as in Eq.~\eqref{eq:parquet} or, if a specific 
channel based notation is chosen, the other channels' arguments have to be projected on the latter.  
This yields for the channel $r$ of the two-particle vertex
\begin{align}
\label{eq:diagrammatic_vertex}
    \mathbf{V}^r = \mathbf{\Lambda}^{\mathrm{2PI}} + \mathbf{\Phi}^r + \sum_{r'\neq r}P^{r' \rightarrow r} \mathbf{\Phi}^{r'} \;.
\end{align}
For the computation of physical observables such as the susceptibilities, it is often more convenient to work directly in 
the physical channels $\eta=\M/\D/\SC$. 
They are related to the diagrammatic channels by
\begin{subequations}
\begin{align}
    \mathbf{\Phi}^{\M} &= -\mathbf{\Phi}^{\xph} \\
    \mathbf{\Phi}^{\D} &= 2\mathbf{\Phi}^{ph} - \mathbf{\Phi}^{\xph} \\
        \mathbf{\Phi}^{\SC} &= \mathbf{\Phi}^{pp} \;.
\end{align}
\end{subequations}
The same relations hold for the components of $\mathbf{V}$. 
Inserting these definitions into Eq.~\eqref{eq:diagrammatic_vertex} gives
\begin{subequations}
\label{eq:vs}
\begin{align}
    \mathbf{V}^\M &= -\mathbf{\Lambda}^{2PI} + \mathbf{\Phi}^\M+ \frac{1}{2}P^{ph \rightarrow \xph} \left(\mathbf{\Phi}^\M - \mathbf{\Phi}^\D \right) -
     P^{pp \rightarrow \xph}\mathbf{\Phi}^{\SC}  \\
    \mathbf{V}^\D &= \mathbf{\Lambda}^{\mathrm{2PI}} - 2P^{\xph \rightarrow ph}\mathbf{\Phi}^\M +\frac{1}{2} P^{ph\rightarrow\xph}\left( \mathbf{\Phi}^\M-\mathbf{\Phi}^\D\right) + \mathbf{\Phi}^\D  - P^{pp \rightarrow \xph}\mathbf{\Phi}^{\SC} + 2P^{pp \rightarrow ph}\mathbf{\Phi}^{\SC}\\
    \mathbf{V}^\SC &= \mathbf{\Lambda}^{\mathrm{2PI}}  - 
    P^{\xph \rightarrow pp} \mathbf{\Phi}^\M - \frac{1}{2}P^{ph \rightarrow pp} \left(\mathbf{\Phi}^\M -  \mathbf{\Phi}^\D \right) + \mathbf{\Phi}^{\SC}\;.
\end{align}
\end{subequations}
Note that while the expressions of $\mathbf{V}^r$
differ from one another only in the notation used for 
their arguments, this is no longer true for $\mathbf{V}^\eta$ in the physical channels. 

\section{Numerical implementation}
\label{app:impl}

\subsection{$2\ell$ flow equations}

The $2\ell$ equation for the flow of the two-particle vertex reads
\begin{align}
    \dot{\mathbf{\Phi}}^{\eta}_{2\ell} = 
    \dot{\mathbf{\Phi}}^{\overline{\eta}}_{1\ell} \mathbf{\Pi}^\eta  \mathbf{\Pi}^\eta +
    \mathbf{V}^\eta \mathbf{\Pi}^\eta \dot{\mathbf{\Phi}}^{\overline{\eta}}_{1\ell} \;,
\end{align}
where $ \mathbf{\Pi}^\eta$ and $ \mathbf{V^\eta}$ are obtained from Eqs. \eqref{eq:pi} and \eqref{eq:vs}, respectively. $\dot{\mathbf{\Phi}}^{\overline{\eta}}_{1\ell}$ represents the sum of the $1\ell$ diagrams of the other channels. The corresponding flow equations read
\begin{subequations}
\begin{align}
    \dot{\mathbf{\Phi}}^{\overline{\M}}_{1\ell} &= 
     -\frac{1}{2}P^{ph\rightarrow \overline{ph}} \left( \dot{\mathbf{\Phi}}^\M_{1\ell} - 
     \dot{\mathbf{\Phi}}^\D_{1\ell} \right) - 
     P^{pp \rightarrow \overline{ph}} \dot{\mathbf{\Phi}}^\SC_{1\ell}\\
    \dot{\mathbf{\Phi}}^{\overline{\D}}_{1\ell} &= 
    - 2P^{\overline{ph}\rightarrow ph} \dot{\mathbf{\Phi}}^\M_{1\ell}
    -\frac{1}{2}P^{ph \rightarrow \overline{ph}} \left( \dot{\mathbf{\Phi}}^\D_{1\ell}-\dot{\mathbf{\Phi}}^\M_{1\ell}\right)
    + 2P^{pp\rightarrow ph}\dot{\mathbf{\Phi}}^\SC_{1\ell} - 
    P^{pp\rightarrow \overline{ph}}\dot{\mathbf{\Phi}}^\SC_{1\ell}  
    \\
    \dot{\mathbf{\Phi}}^{\overline{\SC}}_{1\ell} &= 
    - P^{\xph \rightarrow pp} \dot{\mathbf{\Phi}}_{1\ell}^\M
    + \frac{1}{2}P^{ph \rightarrow pp} \left(\dot{\mathbf{\Phi}}_{1\ell}^\D -
    \dot{\mathbf{\Phi}}_{1\ell}^\M \right)\;,
\end{align}
\end{subequations}
where bold symbols represent matrices in form-factor space.

In order to account for the form-factor truncation, the self-energy flow is determined by the scale derivative of the Schwinger-Dyson equation \cite{Hille2020,Hille2020b} replacing the conventional $1\ell$ flow equation\footnote{We note that the self-energy derivative for the Katanin substitution \cite{Katanin2004} is always computed with the conventional flow, see also Ref.~\citeonline{Hille2020}.}
\begin{align}
    \dot{\Sigma}(k) = \partial_\Lambda \left[ \sum_{k_2,k_3,k_4} V(k,k_2,k_3,k_4)G(k_2)G(k_3)G(k_4)\delta_{k+k_3-k_2,k_4}U\right]\;,
\end{align}
where $G$ is the renormalized propagator including
the self-energy via the Dyson equation $G^{-1}(k)=G_0^{-1}(k)-\Sigma(k)$.  

\subsection{Technical parameters}

We here provide the technical parameters for the 
fRG calculations performed with a smooth frequency cutoff, for the details of the 
algorithmic implementation we refer to Refs. \cite{HeinzelmannThesis}.

We use $n=4$ positive fermionic frequencies that determine the parametrization of the two-particle vertex (see Ref.~\citeonline{HeinzelmannThesis} for the definitions). The rest function contains $(4n+1)\times(2n)\times(2n)$, the $K_2$-function and the fermion-boson vertex $(4n+1)\times(2n)$, the $K_1$-function $(128n+1)$, and the self-energy $(8n)$ frequencies. The fermionic momentum dependence of the vertices and response functions is accounted for by a form-factor expansion, where we considered the local $s$-wave and non-local $d$-wave contributions. The remaining momentum dependence of the vertices, the response functions and the self-energy are calculated on $(16 \times 16)$ equally spaced momentum patches,
with a refinement of $(5 \times 5) $ additional
patches around $\mathbf{q}=(\pi,\pi)$ to resolve the AF peak.
The Green's functions and their summation in the particle-hole and particle-particle bubble are calculated on a $(80 \times 80)$ grid. 

\section{Momentum-resolved susceptibilities} 
\label{app:altfig2}

\begin{figure}
    \centering
    \includegraphics[width=1.\linewidth]{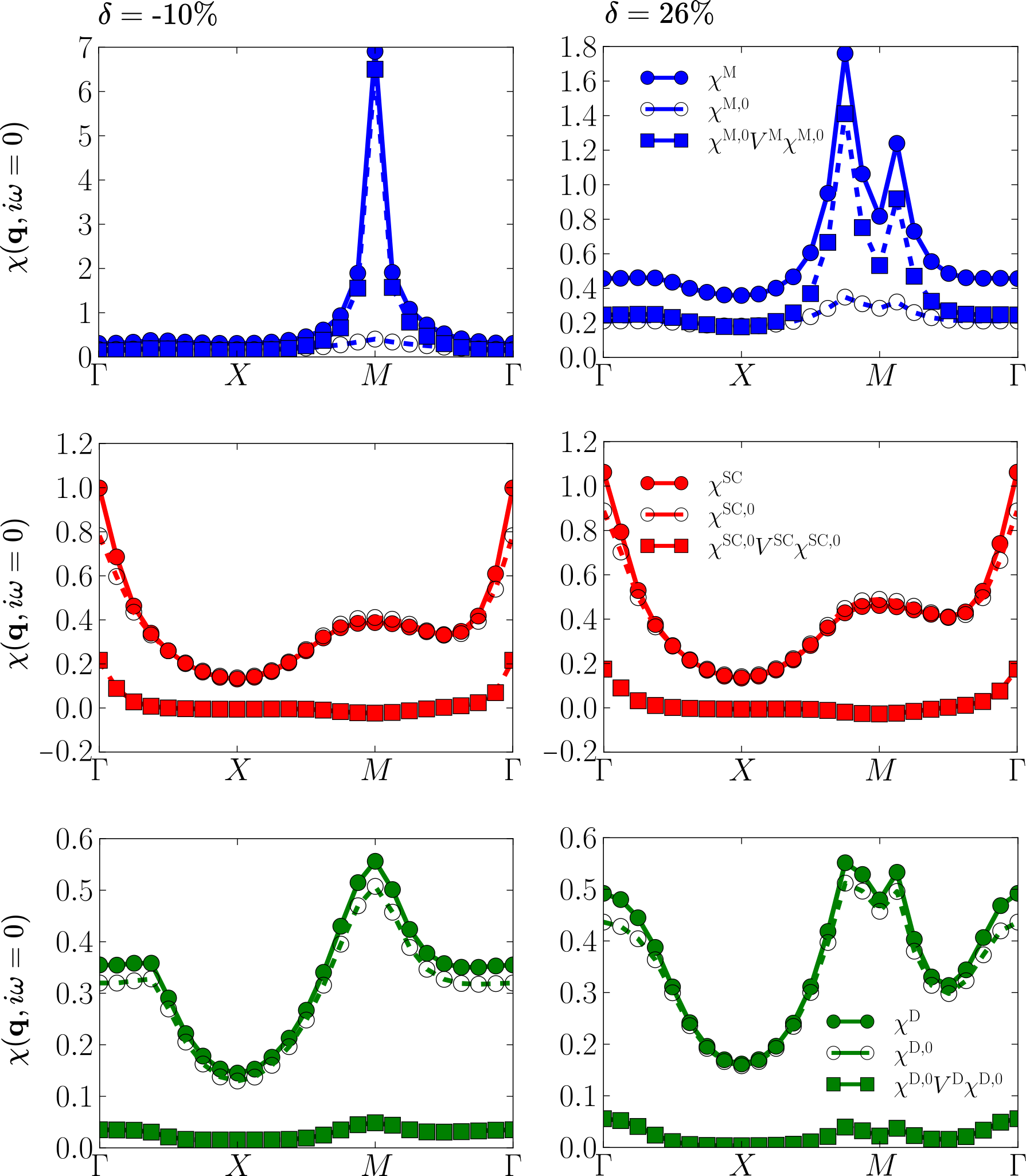}
    \caption{Bubble and vertex contributions to $\chi^{\M/\SC/\D}$ along a path in the Brillouin zone at $10~\%$ electron and $26~\%$ hole doping respectively, for the same parameters as in Fig. \ref{fig:phasediagram_overview}}
    \label{fig:postprocM_alongBZ}
\end{figure}

In Fig. \ref{fig:postprocM_alongBZ} we show the leading magnetic, density, and superconducting susceptibilities of Fig. \ref{fig:susc_alongBZ} along a path in the Brillouin zone, for $\delta=10\%$ electron and $\delta=26\%$ hope doping (left and right panels respectively). 
From the bubble and vertex contributions for the different channels one can infer their importance in driving the respective susceptibilities. In particular, we observe that the maximal values are dominated by the vertex in the magnetic channel, while the density and superconducting ones are determined mainly by the bubble. Nevertheless, a valuable vertex contribution is detected, in particular around half filling.
A more detailed analysis is presented in the sections on the fluctuation diagnostics.

\section{Fermi surface}
\label{app:fermisurface}

\begin{figure}
    \centering
    \includegraphics[width=0.9\linewidth]{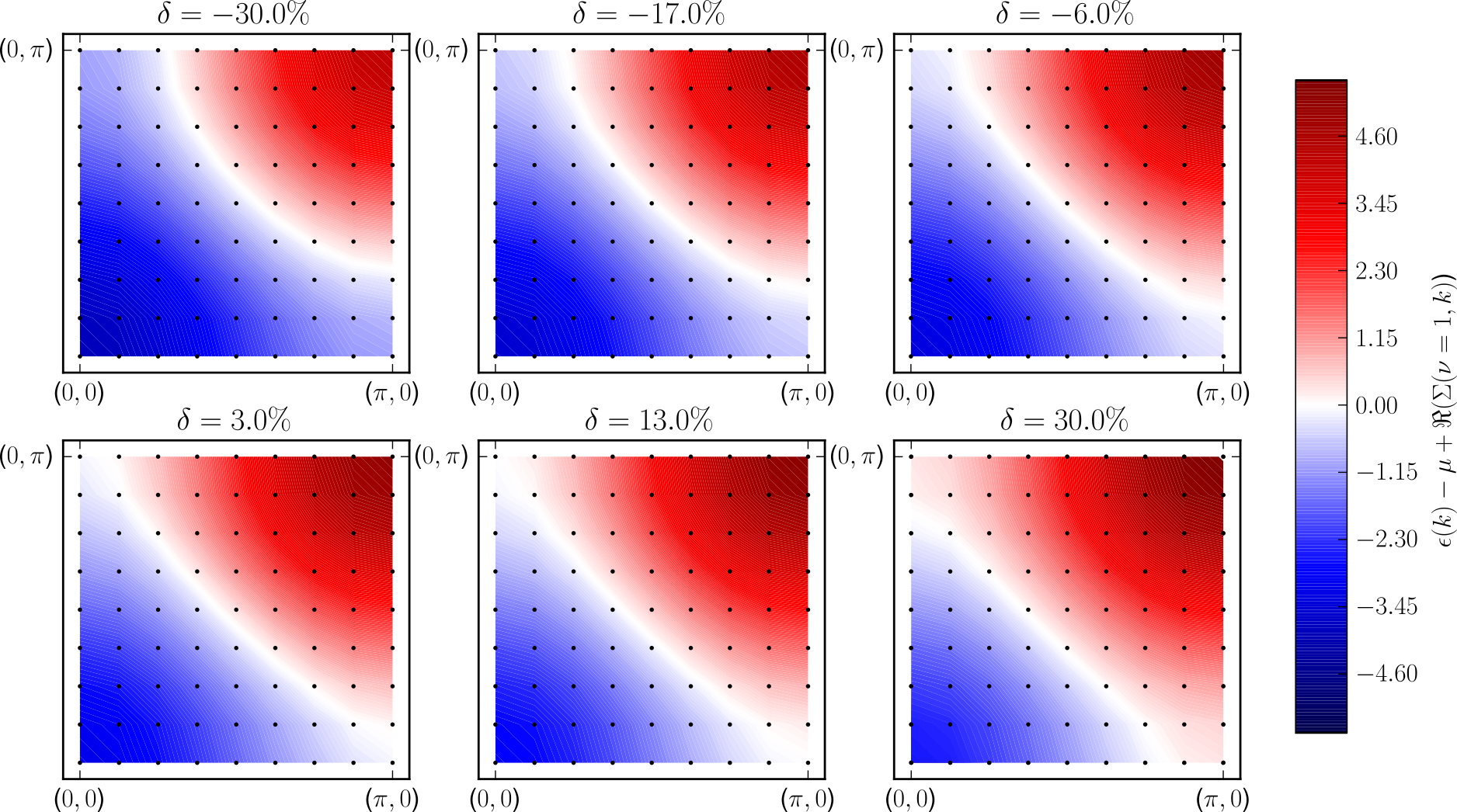}
    \caption{Fermi surfaces corresponding to the suscpetibilities of   
    Fig.~\ref{fig:phasediagram_overview}, for selected values of the doping.}
    \label{fig:fermisurface}
\end{figure}

In Fig.~\ref{fig:fermisurface} we show the Fermi surfaces associated to Fig.~\ref{fig:phasediagram_overview}. For the different values of the doping, the evolution from an electron- to a hole-like shape is clearly visible.

\section{Analytical calculations} 
\label{sec:analyt}

\subsection{Magnetic contribution to the $d$-wave superconducting susceptibility}
\label{sec:AF_SC_connection}
The contribution of the dominant magnetic channel $\mathbf{\Phi}^\M$ to the $d$-wave superconducting susceptibility $\chi^{\SC}_{11}$ (where we indicated the component explicitly) is given by
\begin{align}
    &\frac{3}{2}\left[\chi^{\SC,0}\mathbf{\Phi}^\M\chi^{\SC,0}\right]_{11}(\mathbf{q},i\omega=0) :=  -\sum\limits_{\substack{i\nu,i\nu'  m',n'}}\Pi^{\SC}_{1m'}(\mathbf{q},i\omega=0,i\nu)\nonumber\\
    & \qquad\cdot\left(\frac{1}{2}\left[P^{ph\rightarrow pp}\mathbf{\Phi}^\M\right]_{m'n'} + \left[P^{\xph\rightarrow pp}\mathbf{\Phi}^\M\right]_{m'n'}\right)
    (\mathbf{q},i\omega=0,i\nu,i\nu') 
    \Pi^{\SC}_{n'1}(\mathbf{q},i\omega=0,i\nu') \;.
    \label{eq:appendix_chi_sc}
\end{align}
Since bubbles with mixed form factors are very small and $\Pi^\SC(\mathbf{q}=(0,0),\omega)=0$ exactly, we neglect all terms with $m'=n'=0$.  
Hence, the sum over the different components contains only the elements $n'=m'=1$. 
In the following we drop the frequency dependence to lighten the notation 
(with the sums over internal fermionic frequencies implicitly taken into account).
Projections from one notation to another are necessary because a simple shift of arguments is not possible for the form-factor arguments: 
\begin{align}
    \left[P^{\xph\rightarrow pp}\mathbf{\Phi}^\M\right]_{11} (\mathbf{q}) 
    &= \sum_{m'n'}\int_{\text{BZ}} \text{d} \mathbf{k} \text{d} \mathbf{k'} f^*_{1}(\mathbf{k})f_{1}(\mathbf{k'})f_{m'}(\mathbf{k})f^*_{n'}(\mathbf{k'})\Phi_{m'n'}^{\M,\xph}(\mathbf{q}-\mathbf{k}-\mathbf{k'}) \nonumber\\
    &\!\!\!\!\!\!\!\!\!\!\overset{\mathbf{q}-\mathbf{k}-\mathbf{k'}\rightarrow \mathbf{p}}{=} \sum_{m'n'}\int_{\text{BZ}} \text{d} \mathbf{k} \text{d} \mathbf{p} f^*_{1}(\mathbf{k})f_{1}(\mathbf{q}-\mathbf{k}-\mathbf{p})f_{m'}(\mathbf{k})f^*_{n'}(\mathbf{q}-\mathbf{k}-\mathbf{p})\Phi_{m'n'}^{\M,\xph}(\mathbf{p}) \nonumber\\
    &\approx \int_{\text{BZ}} \text{d} \mathbf{k} \text{d} \mathbf{p} f^*_{1}(\mathbf{k})f_{1}(\mathbf{q}-\mathbf{k}-\mathbf{p})f_{0}(\mathbf{k})f^*_{0}(\mathbf{q}-\mathbf{k}-\mathbf{p})\Phi_{00}^{\M,\xph}(\mathbf{p}) \;,\end{align}
where we used that also contributions from $\Phi^{\M}_{11}$ and mixed components are negligible compared to $\Phi^\M_{00}$. For $\mathbf{q}=0$ and mixed components, the expression vanishes exactly due to the same symmetries as for the mixed bubble. 
Analogously, 
\begin{align}
    \left[P^{ph\rightarrow pp}\mathbf{\Phi}^\M\right]_{11} (\mathbf{q}) &= \sum_{mn}\int_{\text{BZ}} \text{d} \mathbf{k} \text{d} \mathbf{k'} f^*_{1}(\mathbf{k})f_{1}(\mathbf{k'})f_{m}(\mathbf{k})f^*_{n}(\mathbf{q}-\mathbf{k'})\Phi_{mn}^{\M,ph}(\mathbf{k}-\mathbf{k'}) \nonumber\\
    &\approx \int_{\text{BZ}} \text{d} \mathbf{k} \text{d} \mathbf{k'} f^*_{1}(\mathbf{k})f_{1}(\mathbf{k}-\mathbf{p})f_{0}(\mathbf{k})f^*_{0}(\mathbf{k}-\mathbf{p})\Phi_{00}^{\M,ph}(\mathbf{p})\;.
\end{align}
Inserting the above expressions in Eq.~\eqref{eq:appendix_chi_sc} together with the explicit form of the $s$- and  $d$-wave form factors, yields 
\begin{align}
    &\frac{3}{2}\left[\chi^{\SC,0}\mathbf{\Phi}^\M\chi^{\SC,0}\right]_{11}(\mathbf{q}) \nonumber \\
    &=-\Pi^\SC_{11}(\mathbf{q})
    \left(\frac{1}{2} \left[P^{\xph\rightarrow pp}\mathbf{\Phi}^\M\right]_{11} (\mathbf{q}) + \left[P^{ph\rightarrow pp}\mathbf{\Phi}^\M\right]_{11}(\mathbf{q}) \right) \Pi^\SC_{11}(\mathbf{q}) \nonumber \\
    &= -\Pi^\SC_{11}(\mathbf{q}) \left(\frac{1}{2} \int_{\text{BZ}} \text{d} \mathbf{k} \text{d} \mathbf{p} \left(\cos(k_x)-\cos(k_y)\right)\left(\cos(k_x-p_x)-\cos(k_y-p_y)\right)\Phi_{00}^{\M,ph}(\mathbf{p}) \nonumber \right.\\
    &\quad\;\;\;\left.+\int_{\text{BZ}} \text{d} \mathbf{k} \text{d} \mathbf{p} \left(\cos(k_x)-\cos(k_y)\right)\left(\cos(q_x-k_x-p_x)-\cos(q_y-k_y-p_y)\right)\Phi_{00}^{\M,\xph}(\mathbf{p})\right) \Pi^\SC_{11}(\mathbf{q})\nonumber \\    
    &=-\Pi^\SC_{11}(\mathbf{q}) \frac{1}{2}\int_{\text{BZ}} \text{d}\mathbf{p} \left(\frac{1}{2}\left(\cos{p_x}+\cos{p_y}\right)+\cos{(p_x-q_x)}+\cos{(p_y-q_y)}\right) \Phi^{\M}_{00}(\mathbf{p})
    \Pi^\SC_{11}(\mathbf{q})\;,
\end{align}
which reproduces Eq.~\eqref{eq:magnetic_contr_to_chi_sc}. 

\subsection{Lowest order diagram contributing to  $\Phi^{\SC}_{11}$}
\label{app:lowest_order_phiSC_d}
The lowest order vertex diagram from the magnetic channel contributing to the $d$-wave superconducting susceptibility is determined by
\begin{align}
\!\!\!\!\!\!\!\!\!\!\!\!\!\!\!\!\!\!\!\!\!\!\!\!\!\!\!\!\!\!\!\!\!-[P^{\xph\rightarrow pp}\Phi^{\M,\text{lowest order}}_{00}]_{11}(\mathbf{q}) &= U^2 \left[P^{\xph\rightarrow pp}\Pi^{ph}\right]_{11}(\mathbf{q}) \nonumber\\
    &
    = \int_{\text{BZ}} \text{d}\mathbf{k}\text{d}\mathbf{p}f_1(\mathbf{k})f_{1}(\mathbf{p-q+k})\Pi^{ph}_{00}(\mathbf{p})\;.
    \end{align}
With the expressions for the form factors, we obtain    
    \begin{align}
    &
    =\int_{\text{BZ}} \text{d}\mathbf{k}\text{d}\mathbf{p}\left(\cos(k_x)-\cos(k_y)\right)\left(\cos(p_x-q_x+k_x)-\cos(p_y-q_y+k_y)\right)
    \Pi^{ph}_{00}(\mathbf{p}) \nonumber\\
    &=\int_{\text{BZ}} \text{d}\mathbf{p}
    \left(\cos(p_x-q_x)+\cos(p_y-q_y)\right)
    \Pi^{ph}_{00}(\mathbf{p})
\end{align}
and, inserted into $\chi^{\SC,0}\Phi^{\M}_{00}\chi^{\SC,0}$, gives 
\begin{align}
     \chi^{\SC,0}\Phi^{\M,\text{lowest order}}_{00}\chi^{\SC,0}(\mathbf{q}) &= U^2 \left[P^{\xph\rightarrow pp}\Pi^{ph}\right]_{11}(\mathbf{q})\nonumber\\
    &\approx U^2 \;\Pi^\SC_{11}(\mathbf{q})\Pi^\SC_{11}(\mathbf{q})\; \int_{\text{BZ}} \text{d}\mathbf{p}
    \left(\cos(p_x-q_x)+\cos(p_y-q_y)\right)
    \Pi^{ph}_{00}(\mathbf{p})\;,
    \label{eq:lowest_M_in_SCd}
\end{align}
as mixed $pp$-bubbles are very small and vanish exactly for $\mathbf{q}=0$.
This lowest order diagram contribution exhibits the same momentum dependence as the one of Eq.~\eqref{eq:appendix_chi_sc} and illustrates how 
the $s$-wave magnetic channel 
directly affects
$\chi^\SC_{11}$. In particular, we can trace how features at different transfer momenta 
enter 
the maximum of $\chi^\SC_{11}$ at zero transfer momentum, 
the main contribution stemming from $(\pi,\pi)$ 
\begin{align}
    \chi^{\SC,0}\Phi^{\M,\text{lowest order}}_{00}\chi^{\SC,0}(\mathbf{q}=0) = U^2 \;\Pi^\SC_{11}(0)\Pi^\SC_{11}(0)\; \int_{\text{BZ}} \text{d}\mathbf{p}
    \left(\cos(p_x)+\cos(p_y)\right) \Pi^{ph}_{00}(\mathbf{p})\;.
\end{align} 
The trigonometric prefactor in the integral controls the weight of the contributions at different momenta. 
Commensurate AF features at $\mathbf{p}=(\pi,\pi)$ are most favoured by the projection with a weight of $-1$, whereas incommensurate ones 
are reduced by a factor of $(\cos(p_x-\delta_x)+\cos(p_y-\delta_y))$. Ferromagnetic features at $\mathbf{p}=0$ have the same weight as AF contributions, but with opposite sign with an overall suppressing effect \cite{Dong2022}. 

Likewise, we obtain for 
\begin{align}
\!\!\!\!\!\!\!\!\!\!\!\!\!\!\!\!\!\!\!\!\!\!\!\!\!\!\!\!\!\!\!\!\!\!\!\!\!\!\!\!\!\!\!\!\!\Phi^{\SC,\text{lowest order}}_{11}(\mathbf{q}) & =U^4\sum_{m'n'} \; \left[P^{\xph\rightarrow pp}\Pi^{ph}\right]_{1m'}(\mathbf{q})\;\Pi^\SC_{m'n'}(\mathbf{q})\;\left[P^{\xph\rightarrow pp}\Pi^{ph}\right]_{n'1}(\mathbf{q}) \;, 
  \end{align}
  with
    \begin{align} 
    &=U^4\sum_{m'n'}\left(\int_{\text{BZ}} \text{d}\mathbf{k}\text{d}\mathbf{p}f_1(\mathbf{k})f_{m'}(\mathbf{p-q+k})\Pi^{ph}_{00}(\mathbf{p})\right)\Pi^\SC_{m'n'}(\mathbf{q}) \nonumber\\
    &\qquad\;\;\left(\int_{\text{BZ}} \text{d}\mathbf{k'}\text{d}\mathbf{p'}f_1(\mathbf{k'})f_{n'}(\mathbf{p'-q+k'})\Pi^{ph}_{00}(\mathbf{p'})\right) \nonumber\\
    &=U^4\left(\int_{\text{BZ}} \text{d}\mathbf{k}\text{d}\mathbf{p}\left(\cos(k_x)-\cos(k_y)\right)\left(\cos(p_x-q_x+k_x)-\cos(p_y-q_y+k_y)\right)\Pi^{ph}_{00}(\mathbf{p})\right) \Pi^\SC_{11}(\mathbf{q})\nonumber\\
    &\qquad\;\;\left(\int_{\text{BZ}} \text{d}\mathbf{k'}\text{d}\mathbf{p'}\left(\cos(k_x)-\cos(k_y)\right)\left(\cos(p'_x-q_x+k_x)-\cos(p'_y-q_y+k_y)\right)\Pi^{ph}_{00}(\mathbf{p'})\right) \nonumber\\  
    &=U^4\left(\int_{\text{BZ}} \text{d}\mathbf{p}\left(\cos(p_x-q_x)+\cos(p_y-q_y)\right)\Pi^{ph}_{00}(\mathbf{p})\right) \Pi^\SC_{11}(\mathbf{q})\nonumber\\
    &\qquad\;\;\left(\int_{\text{BZ}} \text{d}\mathbf{p'}\left(\cos(p'_x-q_x)+\cos(p'_y-q_y)\right)\Pi^{ph}_{00}(\mathbf{p'})\right)\;,
\end{align}
where 
we used the fact that the integral over $\mathbf{k}/\mathbf{k'}$ vanishes for $m'/n'=0$ and hence
\begin{align}
    \chi^{\SC,0}\Phi^{\SC,\text{lowest order}}_{11}\chi^{\SC,0} (\mathbf{q})& 
    \nonumber\\
    &\!\!\!\!\!\!\!\!\!\!\!\!\!\!\!\!\!\!\!\!\!\!\!\!\!\!\!\!\!\!=U^4\Pi^\SC_{11}(\mathbf{q})\left(\int_{\text{BZ}} \text{d}\mathbf{p}\left(\cos(p_x-q_x)+\cos(p_y-q_y)\right)\Pi^{ph}_{00}(\mathbf{p})\right) \Pi^\SC_{11}(\mathbf{q})\nonumber\\
    &\!\!\!\!\!\!\!\!\!\!\!\!\!\!\!\!\!\!\!\!\!\!\!\!\!\!\!\!\!\!\qquad\qquad\quad\;\;\,\left(\int_{\text{BZ}} \text{d}\mathbf{p'}\left(\cos(p'_x-q_x)+\cos(p'_y-q_y)\right)\Pi^{ph}_{00}(\mathbf{p'})\right) \Pi^\SC_{11}(\mathbf{q})\;.
\end{align}
In fact, these represent the lowest order contributions in the $pp$-ladder, with the insertion of $\left[P^{\xph\rightarrow pp}\Pi^{ph}\right]_{11}$ instead of the bare (local) interaction, as illustrated in \cfg{xph_in_chiSC_d}. 

\begin{figure}
    \centering
    \includegraphics[width=0.9\linewidth]{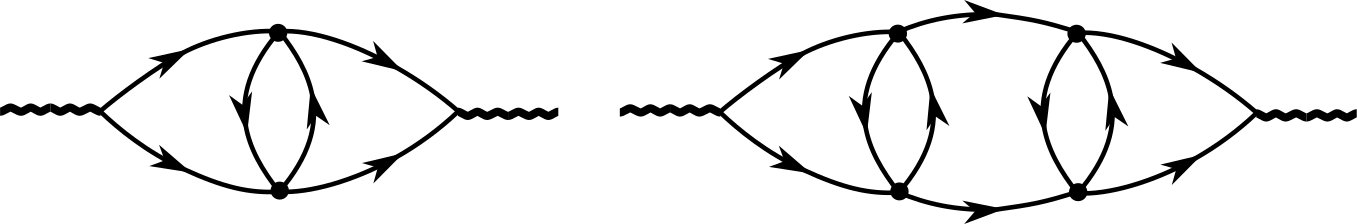}
    \caption{First and second order diagrams of a ladder expansion in the $pp$-channel with $\left[P^{\xph\rightarrow pp}\Pi^{ph}\right]_{11}$ instead of the bare interaction.}
    \label{fig:xph_in_chiSC_d}
\end{figure}

\end{appendix}

\bibliography{SciPost_Example_BiBTeX_File.bib}

\nolinenumbers

\end{document}